\documentclass[apj]{emulateapj}
\usepackage{mathptmx}
\usepackage{lscape}
\usepackage{rotate}

\newcommand{\ltsim}{\raisebox{-.5ex}{$\;\stackrel{<}{\sim}\;$}}
\newcommand{\gtsim}{\raisebox{-.5ex}{$\;\stackrel{>}{\sim}\;$}}
\newcommand{\kms}{\ifmmode {\rm km\ s}^{-1} \else km s$^{-1}$\fi}
\newcommand{\msun}{$M_{\odot}$}
\newcommand{\et}{et al.\ }
\newcommand{\xray}{\hbox{X-ray}}
\newcommand{\aox}{$\alpha_{\rm ox}$}
\newcommand{\aro}{$\alpha_{\rm ro}$}
\newcommand{\daox}{$\Delta\alpha_{\rm ox}$}
\newcommand{\nh}{$N_{\rm H}$}
\newcommand{\mb}{$M_{\rm B}$}
\newcommand{\Ka}{Fe K$\alpha$}
\newcommand{\xmm}{{\sl XMM-Newton}}
\newcommand{\chandra}{{\sl Chandra}}

\journalinfo{The Astrophysical Journal, ???:??--??, 2006 ???,
  astro-ph/0602442}
\slugcomment{Received 2006 January 6; accepted 2006 February 17}

\shorttitle{{\slshape {CHANDRA}} OBSERVATIONS OF
  HIGH-REDSHIFT SDSS QUASARS}
\shortauthors{SHEMMER ET AL.}

\begin{document}

\title{{\slshape {Chandra}} OBSERVATIONS OF THE HIGHEST
  REDSHIFT QUASARS FROM THE SLOAN DIGITAL SKY SURVEY}

\author{
Ohad Shemmer,\altaffilmark{1}
W.\,N. Brandt,\altaffilmark{1}
Donald P. Schneider,\altaffilmark{1}
Xiaohui Fan,\altaffilmark{2}
Michael A. Strauss,\altaffilmark{3}
Aleksandar~M.~Diamond-Stanic,\altaffilmark{2}
Gordon T. Richards,\altaffilmark{3}
Scott F. Anderson,\altaffilmark{4}
James E. Gunn,\altaffilmark{3} \\
and Jon Brinkmann\altaffilmark{5}
}

\altaffiltext{1}
               {Department of Astronomy \& Astrophysics, The Pennsylvania
               State University, University Park, PA 16802;
               ohad@astro.psu.edu}
\altaffiltext{2}
               {Steward Observatory, University of Arizona, 933 North Cherry
               Avenue, Tucson, AZ 85721}

\altaffiltext{3}
               {Princeton University Observatory, Peyton Hall, Princeton,
               NJ 08544}

\altaffiltext{4}
               {Department of Astronomy, University of Washington, Box 351580,
               Seattle, WA 98195}

\altaffiltext{5}
               {Apache Point Observatory, P.O. Box 59, Sunspot,
               NM 88349-0059}

\begin{abstract}
We present new \chandra\ observations of 21 $z>4$ quasars, including
11 sources at $z>5$. These observations double the number of
\xray\ detected quasars at $z>5$, allowing investigation of the
\xray\ spectral properties of a substantial sample of quasars at the
dawn of the modern Universe. By jointly fitting the spectra of 15
$z>5$ radio-quiet quasars (RQQs), including sources from the
\chandra\ archive, with a total of 185 photons, we find a mean
\xray\ power-law photon index of $\Gamma=1.95^{+0.30}_{-0.26}$, and a
mean neutral intrinsic absorption column density of
\nh\ltsim6$\times10^{22}$\,cm$^{-2}$. These results show that quasar
\xray\ spectral properties have not evolved up to the highest
observable redshifts. We also find that the mean
optical--\xray\ spectral slope (\aox) of optically-selected $z>5$
RQQs, excluding broad absorption line quasars, is \aox\,$=-1.69\pm0.03$,
which is consistent with the value predicted from the observed
relationship between \aox\ and ultraviolet luminosity. Four of the
sources in our sample are members of the rare class of weak
emission-line quasars, and we detect two of them in \hbox{X-rays}. We
discuss the implications our \xray\ observations have for the nature
of these mysterious sources and, in particular, whether their
weak-line spectra are a consequence of continuum boosting or a deficit
of high-ionization line emitting gas.
\end{abstract}


\section{Introduction}
\label{introduction}

During the past five years over 100 quasars at $z>4$ have been
detected in X-rays\footnote{See
  http://www.astro.psu.edu/users/niel/papers/highz-xray-detected.dat
  for a regularly updated compilation of \xray\ detections at
  \hbox{$z>4$}.}, probing the inner regions of active galactic nuclei
(AGNs) when the Universe was \ltsim1.5\,Gyr old (e.g., Kaspi \et 2000;
Brandt \et 2002; Bechtold \et 2003; Vignali \et 2003a, 2003b, 2005;
Shemmer \et 2005). Those studies have provided measurements of the
\xray\ power-law photon index, intrinsic absorption column density,
and the optical--X-ray spectral energy distribution (SED) of a few
$z>4$ quasars which are particularly bright in \hbox{X-rays}, as well
as the mean properties of large samples of these sources via joint
fitting of their \xray\ spectra (e.g., Page \et 2005; Shemmer \et
2005; Vignali \et 2003a, 2003b, 2005). The results demonstrated that
the \xray\ spectra of $z>4$ quasars do not differ significantly from
those of AGNs at lower redshift, and that the \xray\ photon index of
AGNs does not depend on luminosity (but see also Bechtold \et 2003;
Dai \et 2004; Grupe \et 2006). This is generally consistent with
recent observations showing that the longer wavelength SEDs of quasars
have not significantly evolved over cosmic time (e.g., see Stern \et
2000, Carilli \et 2001, and Petric \et 2003 for radio observations and
Schneider, Schmidt, \& Gunn 1991, Vanden Berk \et 2001, and Pentericci
\et 2003 for rest-frame \hbox{optical--ultraviolet (UV)}
observations). While the \xray\ spectral properties of $4<z<5$ quasars
are now fairly well established, the situation for quasars at the
highest observable redshifts, $z>5$, is not yet as clear. Only loose
constraints on the typical \xray\ spectral properties of $z>5$ quasars
are available, mainly due to the small number of such sources
discovered so far (\ltsim50). Obtaining \xray\ spectral information
for the highest-redshift quasars is important for understanding the
physical processes and environments of AGNs at the end of the
reionization epoch (e.g., Loeb \& Barkana 2001).

\begin{figure*}
\plotone{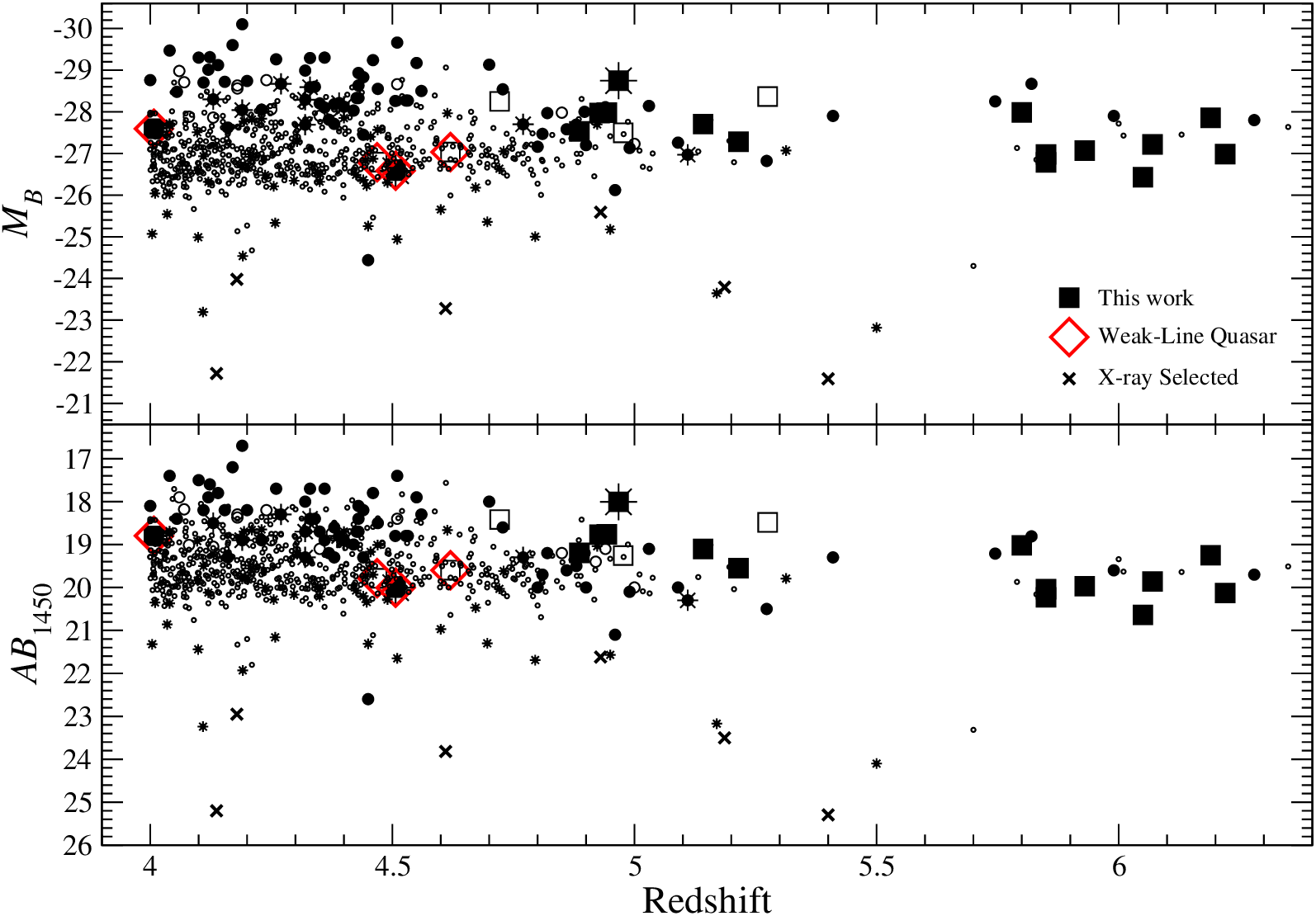}
\caption{\mb\ ({\it top}) and $AB_{1450}$ ({\it bottom}) vs. redshift
  for all published $z>4$ quasars (data were compiled from this work,
  Bassett \et 2004, Steffen \et 2004, Mahabal \et 2005, Vignali \et
  2005, and Fan \et 2006). Large symbols indicate \xray\ targeted
  quasars. Open symbols indicate targeted quasars that are
  \xray-undetected, but have meaningful upper limits (i.e.,
  \aox\,$<-1.50$). Stars indicate RLQs and radio-moderate
  quasars. Squares indicate sources from this work, and enclosing
  diamonds indicate the WLQs in our sample. Note that two other $z>4$
  WLQs, SDSS\,J0040$-$0915 (Schneider \et 2003) and
  SDSS\,J133550.81$+$353315.8 (Fan \et 2006), are not marked with an
  enclosing diamond, for clarity.}
\label{zgt4}
\end{figure*}

X-ray observations of $z>5$ quasars are challenging for present-day
\xray\ missions such as the \chandra\ \xray\ Observatory and
\xmm. Typically, exposure times of $\approx10$\,ks or more are
required to detect $\sim5$ photons from these sources, and
\gtsim100\,ks is required to obtain a reliable measurement of the
photon index (Farrah \et 2004; Schwartz \& Virani 2004). Prior to this
work, only 11 quasars (10 of which are radio quiet, and one that is
radio loud) at $z>5$ had been \xray\ detected (Brandt \et 2001, 2002;
Vignali \et 2001, 2002, 2003b; Barger \et 2002; Mathur \et 2002;
Bechtold \et 2003; Steffen \et 2004). In this work we add 11 more
\xray\ detections of $z>5$ quasars\footnote{The \xray\ detections of
  two of these sources, SDSS\,J1053$+$5804 and SDSS\,J2228$-$0757,
  were reported briefly by Schneider \et (2005); here we provide a
  full description of~their~detections~(see \S\,\ref{results}).} and
create a statistically meaningful sample to investigate the mean
\xray\ properties of these sources. In addition, we present new
\xray\ observations of four $z\sim4.5$ sources which belong to the
rare class of weak (or absent) emission-line quasars (hereafter WLQs);
only about a dozen sources of this type have been discovered at $z>2$
(e.g., Fan \et 1999; Anderson \et 2001; Collinge \et 2005; Schneider
\et 2005).

This paper is organized as follows: in \S\,\ref{observations} we
describe the \chandra\ observations and their reduction; the results
are presented in \S\,\ref{results}. In \S\,\ref{individual} we provide
notes on individual interesting sources in our sample. In
\S\,\ref{xray_properties} we discuss the mean \xray\ spectral
properties of $z>5$ quasars. In \S\,\ref{WLQ} we discuss the results
of our \xray\ observations of WLQs, and how they may shed light on the
nature of these sources. A summary of our main results is given in
\S\,\ref{summary}. Complete quasar names are given in the tables and
in the headers of \S\,\ref{individual}, and their abbreviated versions
are used throughout the text. Luminosity distances in the paper are
determined using the standard cosmological model with parameters
$\Omega_{\Lambda}=0.7$, $\Omega_{M}=0.3$, and
\hbox{$H_{0}=70$\,\kms\ Mpc$^{-1}$} (e.g., Spergel \et 2003).

\section{Sample Selection, Observations, and Data Reduction}
\label{observations}

We have targeted some of the optically brightest quasars at $z$\gtsim5
(see Fig.\,\ref{zgt4}) to investigate quasar \xray\ spectral
properties and optical--X-ray SEDs at the highest observable
redshifts. Targets with $z$\ltsim5.4 were mostly selected from the
Sloan Digital Sky Survey (SDSS; York \et 2000) Data Release~3 (DR3)
quasar catalog (Schneider \et 2005). Quasars at higher redshifts
cannot be detected with the SDSS quasar-selection algorithm (Richards
\et 2002), since Ly$\alpha$ shifts from the $i$ band to the $z$ band,
and detections require the aid of near-infrared (IR) imaging and
spectroscopy (e.g., Fan \et 2004). We obtained short
(\hbox{$\sim$3--30\,ks}) \xray\ observations of 19 of the most distant
quasars known with \chandra\ during Cycles 4 and 6 (see
Fig.\,\ref{zgt4}). The observation log appears in Table\,\ref{log},
where we also list references to the quasars' discovery papers. The
targets include 18 newly discovered quasars from the SDSS survey (12
of which appear in the DR3 catalog; Schneider \et 2005) and one
previously \xray-targeted quasar, SDSS\,J1532$-$0039 at $z=4.62$,
which is the prototype of the WLQ class (Fan \et 1999; Vignali \et
2001). Three other WLQs from DR3 are included in our sample; our aim
is to utilize the \xray\ data to shed light on the nature of these
sources. Our sample includes one radio-loud quasar (RLQ),
SDSS\,J0011$+$1446, and two broad absorption line quasars (BALQSOs),
SDSS\,J1048$+$4637 and SDSS\,J1653$+$4054. The Cycle 4 and 6 sources
were observed with the Advanced CCD Imaging Spectrometer (ACIS;
Garmire \et 2003) with the S3 CCD at the aimpoint. In addition, we add
to our sample serendipitous \chandra\ detections of two $z>4$ quasars,
SDSS\,J1026$+$4719 at $z=4.94$ from Cycle~1, and SDSS\,J1053$+$5804 at
$z=5.21$ from Cycle~2 (see Table\,\ref{log}); both of these sources
were observed with ACIS-I and lie near the edge of the field of view.

\begin{deluxetable*}{lcclcccl}
\tablecolumns{8} 
\tabletypesize{\scriptsize}
\tablewidth{0pc}
\tablecaption{\chandra\ Observations Log}
\tablehead{
\colhead{Object} &
\colhead{} &
\colhead{$\Delta_{\rm Opt-X}$\tablenotemark{a}} & 
\colhead{\xray} &
\colhead{\chandra} &
\colhead{Exp. Time\tablenotemark{b}} &
\colhead{} &
\colhead{} \\
\colhead{(SDSS\,J)} &
\colhead{$z$} &
\colhead{(arcsec)} &
\colhead{Obs. Date} & 
\colhead{Cycle no.} &
\colhead{(ks)} &
\colhead{Ref.} &
\colhead{Notes}
}
\startdata
000239.39$+$255034.8 & 5.80 & 0.3 & 2005 Jan 24 & 6 & 5.87 & 1 & \\
000552.34$-$000655.8 & 5.85 & 0.4 & 2005 Jul 28 & 6 & 16.93& 1 & \\
001115.23$+$144601.8 & 4.97 & 0.3 & 2003 May 20 & 4 & 3.49 & 2 & \\
084035.09$+$562419.9 & 5.85 & 0.2 & 2005 Feb 3--4 & 6 & 15.84& 3 & \\
102622.89$+$471907.0 & 4.94 & 1.6 & 2000 June 7--8 & 1 &
2.29\tablenotemark{e} & 2 & Off-axis on ACIS-I \\
104845.05$+$463718.3 & 6.19 & 0.3 & 2005 Jan 10 & 6 & 14.97& 4 & Broad
absorption line quasar \\
105322.98$+$580412.1 & 5.21 & 0.8 & 2001 July 5 & 2 &
4.70\tablenotemark{e} & 2 & Off-axis on ACIS-I \\ \\
130216.13$+$003032.1 & 4.47 & \nodata & 2003 May 17 & 4 & 10.81& 2 &
Weak emission lines \\
140850.91$+$020522.7 & 4.01 & 0.5 & 2005 Mar 02 & 6 & 6.00 & 2 & Weak
emission lines \\
141111.29$+$121737.4 & 5.93 & 0.2 & 2005 Mar 17 & 6 & 14.27& 1 & \\
143352.20$+$022713.9 & 4.72 & \nodata & 2003 Apr 20 & 4 & 3.50 & 2 &
\\
144231.72$+$011055.2 & 4.51 & 0.3 & 2002 Dec 16 & 4 & 11.01& 2 & Weak
emission lines \\
153259.96$-$003944.1\tablenotemark{c} & 4.62 & \nodata & 2005 Apr 2 &
6 & 9.83 & 5 & Weak emission lines \\
153650.26$+$500810.3 & 4.93 & 0.2 & 2005 Sep 23 & 6 & 4.62 & 2 & \\ \\
160253.98$+$422824.9\tablenotemark{d} & 6.07 & 0.3 & 2005 Oct 29 & 6 &
13.20 & 1 & \\
162331.81$+$311200.5 & 6.22 & 0.5 & 2004 Dec 29 & 6 & 17.21& 1 &
Strong Ly$\alpha$ \\
162626.50$+$275132.4 & 5.28 & \nodata & 2005 May 12--13& 6 & 4.82 & 2
& \\
163033.90$+$401209.6 & 6.05 & 0.1 & 2005 Nov 4 & 6 & 27.39 & 4 & \\
165354.61$+$405402.1 & 4.98 & \nodata & 2002 Dec 8--9 & 4 & 3.12 & 2 &
Broad absorption line quasar \\
222509.19$-$001406.8 & 4.89 & 0.4 & 2003 Sep 10--11 & 4 & 3.45 & 2 &
\\
222845.14$-$075755.3 & 5.14 & 0.7 & 2003 May 5 & 4 & 7.04 & 2 &
\enddata
\tablecomments{The optical positions of the quasars have been obtained
from the reference given in the seventh column, and the \xray\
positions have been obtained with {\sc wavdetect}.}
\tablenotetext{a}{Distance between the optical and \xray\ positions;
  missing values indicate no \xray\ detection.}
\tablenotetext{b}{The \chandra\ exposure time has been corrected for
  detector dead time.}
\tablenotetext{c}{An additional 5.09\,ks \chandra\ exposure of the
  source is included in the current analysis, raising the total
  exposure time to 14.92\,ks (see Vignali \et 2001 and
  \S\,\ref{observations} for details).}
\tablenotetext{d}{The optical coordinates of this source from Fan \et
  (2004) have been corrected using the SDSS DR4 astrometry.}
\tablenotetext{e}{Actual exposure time, not corrected by the exposure
  map.}
\tablerefs{(1) Fan \et (2004); (2) Schneider \et (2005); (3) Fan \et
  (2006); (4) Fan \et (2003); (5) Fan \et (1999).}
\label{log}
\end{deluxetable*}

Faint mode was used for the event telemetry format in all the
observations, and {\sl ASCA} grade 0, 2, 3, 4, and 6 events were used
in the analysis, which was carried out using standard {\sc
  ciao\footnote{\chandra\ Interactive Analysis of Observations. See
    http://asc.harvard.edu/ciao/} v3.2} routines. No background flares
are present in these observations. Source detection was carried out
with {\sc wavdetect} (Freeman \et 2002) using wavelet transforms (with
wavelet scale sizes of 1, 1.4, 2, 2.8, and 4 pixels) and a
false-positive probability threshold of 10$^{-4}$. Given the small
number of pixels being searched due to the accurate a priori source
positions and the subarcsecond on-axis angular resolution of \chandra,
the probability of spurious detections is extremely low; most of the
sources were in fact detected at a false-positive probability
threshold of 10$^{-6}$.

The \xray\ counts detected in the ultrasoft band
(\hbox{0.3--0.5\,keV}), the soft band (\hbox{0.5--2\,keV}), the hard
band (\hbox{2--8\,keV}), and the full band (\hbox{0.5--8\,keV}) are
reported in Table\,\ref{counts}. The counts were derived from manual
aperture photometry, and they are consistent with the photometry
obtained using {\sc wavdetect}. Sixteen quasars were detected with
\hbox{2--127} full-band counts; the largest number of counts was
obtained from SDSS\,J0011$+$1446 which is a RLQ. The \xray\ positions
of the 14 targeted and detected sources differ by
\hbox{0.2\arcsec--0.7\arcsec} from their optical positions; the
\xray\ positions of the two serendipitous, off-axis detections differ
by \hbox{0.8\arcsec--1.6\arcsec} from their optical positions
(Table\,\ref{log}). These results are consistent with the expected
\chandra\ ACIS positional errors. We used the {\sc wavdetect} tool a
second time on the fields of the five undetected sources, with a
false-positive probability threshold of 10$^{-3}$, but none of the
five sources was detected even at this lower significance
level. Table\,\ref{counts} also includes the band ratio, calculated as
the counts in the hard band divided by the counts in the soft band,
and the effective power-law photon index ($\Gamma$, where $N(E)\propto
E^{-\Gamma}$; $\Gamma$ was calculated from the band ratio using the
\chandra\ {\sc pimms} v3.6a tool\footnote{See
  http://cxc.harvard.edu/toolkit/pimms.jsp}) for each source. For the
Cycle 4 and 6 sources the photon index was corrected for the
quantum-efficiency decay of ACIS at low energies, caused by molecular
contamination of the ACIS filters, using a time-dependent correction
calculated with {\sc pimms}.

\begin{deluxetable*}{lcccccc}
\tablecolumns{7}
\tabletypesize{\scriptsize}
\tablewidth{0pt}
\tablecaption{X-ray Counts, Band Ratios, and Effective Photon Indices}
\tablehead{ 
\colhead{Object} &
\multicolumn{4}{c}{X-ray Counts\tablenotemark{a}} \\
\cline{2-5} \\
\colhead{(SDSS\,J)} &
\colhead{0.3--0.5\,keV} &
\colhead{0.5--2\,keV} & 
\colhead{2--8\,keV} &
\colhead{0.5--8\,keV} & 
\colhead{Band Ratio\tablenotemark{b}} &
\colhead{$\Gamma$\tablenotemark{b}}
}
\startdata
000239.39$+$255034.8 & $<4.8$ & {\phn}5.0$^{+3.4}_{-2.2}$ & $<3.0$ &
{\phn}5.0$^{+3.4}_{-2.2}$ & $<0.60$ & $>1.2$ \\
000552.34$-$000655.8 & $<3.0$ & {\phn}16.4$^{+5.1}_{-4.0}$ & $<3.0$ &
{\phn}17.6$^{+5.3}_{-4.1}$ & $<0.18$ & $>2.2$ \\
001115.23$+$144601.8 & {\phn}6.9$^{+3.8}_{-2.6}$
&{\phn}98.1$^{+10.9}_{-9.9}$ & {\phn}28.8$^{+6.4}_{-5.3}$ &
{\phn}126.9$^{+12.3}_{-11.2}$ & {\phn}0.29$^{+0.07}_{-0.06}$ &
{\phn}1.7$^{+0.2}_{-0.2}$ \\
084035.09$+$562419.9 & $<3.0$ & {\phn}3.0$^{+2.9}_{-1.6}$ & $<4.8$ &
{\phn}3.0$^{+2.9}_{-1.6}$ & $<1.60$ & $>0.3$ \\
102622.89$+$471907.0 & {\phn}$<3.0$ & {\phn}3.7$^{+3.1}_{-1.8}$ &
{\phn}$<6.4$ & {\phn}5.3$^{+3.5}_{-2.2}$ & {\phn}$<1.73$ &
{\phn}$>0.2$ \\
104845.05$+$463718.3 & $<3.0$ & {\phn}3.0$^{+2.9}_{-1.6}$ & $<3.0$ &
{\phn}3.0$^{+2.9}_{-1.6}$ & $<1.00$ & $>0.7$ \\
105322.98$+$580412.1 & {\phn}$<3.0$ &{\phn}3.7$^{+3.1}_{-1.8}$ &
{\phn}$<4.8$ & {\phn}4.3$^{+3.3}_{-2.0}$ & {\phn}$<1.30$ &
{\phn}$>0.5$ \\ \\
130216.13$+$003032.1 & $<3.0$ & $<4.8$ & $<4.8$ & $<6.4$ & \nodata &
\nodata \\
140850.91$+$020522.7 & {\phn}4.0$^{+3.2}_{-1.9}$ &
26.2$^{+6.2}_{-5.1}$ & {\phn}4.0$^{+3.2}_{-1.9}$ &
30.0$^{+6.5}_{-5.4}$ & {\phn}0.15$^{+0.13}_{-0.08}$ &
{\phn}2.4$^{+0.7}_{-0.6}$ \\
141111.29$+$121737.4 & $<4.8$ & {\phn}10.9$^{+4.4}_{-3.3}$ & $<6.4$ &
{\phn}12.9$^{+4.7}_{-3.5}$ & $<0.59$ & $>1.2$ \\
143352.20$+$022713.9 & $<3.0$ & $<4.8$ & $<3.0$ & $<4.8$ & \nodata &
\nodata \\
144231.72$+$011055.2 & {\phn}2.0$^{+2.7}_{-1.3}$ &
{\phn}36.7$^{+7.1}_{-6.0}$ & {\phn}6.9$^{+3.8}_{-2.6}$ &
{\phn}43.6$^{+7.7}_{-6.6}$ & {\phn}0.19$^{+0.11}_{-0.08}$ &
{\phn}2.0$^{+0.4}_{-0.4}$ \\
153259.96$-$003944.1 & $<3.0$ & $<3.0$ & $<3.0$ & $<3.0$ & \nodata &
\nodata \\
153650.26$+$500810.3 & $<$3.0 & {\phn}7.0$^{+3.8}_{-2.6}$ &
{\phn}8.0$^{+4.0}_{-2.8}$ & 14.7$^{+4.9}_{-3.8}$ &
{\phn}1.15$^{+0.70}_{-0.78}$ & {\phn}0.6$^{+1.0}_{-0.5}$ \\ \\
160253.98$+$422824.9 & $<$4.8 & {\phn}22.6$^{+5.8}_{-4.7}$ &
{\phn}2.9$^{+2.9}_{-1.6}$ & {\phn}26.3$^{+6.2}_{-5.1}$ &
{\phn}0.13$^{+0.13}_{-0.08}$ & {\phn}2.5$^{+0.8}_{-0.6}$ \\
162331.81$+$311200.5 & $<3.0$ & {\phn}2.9$^{+2.9}_{-1.6}$ & $<8.0$ &
{\phn}5.7$^{+3.5}_{-2.3}$ & $<2.76$ & $>-0.2$ \\
162626.50$+$275132.4 & $<3.0$ & $<4.8$ & $<3.0$ & $<4.8$ & \nodata &
\nodata \\
163033.90$+$401209.6 & 2.0$^{+2.7}_{-1.3}$ & 13.6$^{+4.8}_{-3.6}$ &
4.0$^{+3.2}_{-1.9}$ & 17.4$^{+5.3}_{-4.1}$ &
{\phn}0.29$^{+0.25}_{-0.18}$ & {\phn}1.8$^{+0.8}_{-0.6}$ \\
165354.61$+$405402.1 & $<3.0$ & $<3.0$ & $<6.4$ & $<6.4$ & \nodata &
\nodata \\
222509.19$-$001406.8 & $<3.0$ & {\phn}2.0$^{+2.7}_{-1.3}$ & $<3.0$ &
{\phn}2.0$^{+2.7}_{-1.3}$ & $<1.50$ & $>0.2$ \\
222845.14$-$075755.3 & $<3.0$ & {\phn}2.0$^{+2.7}_{-1.3}$ & $<4.8$ &
{\phn}3.0$^{+2.9}_{-1.6}$ & $<2.40$ & $>-0.2$
\enddata
\tablenotetext{a}{Errors on the \xray\ counts were computed according
  to Tables~1 and 2 of Gehrels (1986) and correspond to the 1$\sigma$
  level; these were calculated using Poisson statistics.  The upper
  limits are at the 95\% confidence level and were computed according
  to Kraft, Burrows, \& Nousek (1991). Upper limits of 3.0, 4.8, 6.4,
  and 8.0 indicate that 0, 1, 2, and 3 \xray\ counts, respectively,
  have been found within an extraction region of radius 1\arcsec\
  centered on the optical position of the quasar (considering the
  background within this source-extraction region to be negligible).}
\tablenotetext{b}{We calculated errors at the 1$\sigma$ level for the
  band ratio (the ratio between the \hbox{2--8\,keV} and
  \hbox{0.5--2\,keV} bands) and effective photon index following the
  ``numerical method'' described in $\S$\,1.7.3 of Lyons (1991); this
  avoids the failure of the standard approximate-variance formula when
  the number of counts is small (see $\S$\,2.4.5 of Eadie et
  al. 1971). The photon indices have been obtained applying the
  correction required to account for the ACIS quantum-efficiency decay
  at low energy.}
\label{counts}
\end{deluxetable*}

Figure\,\ref{images1} presents the full-band \chandra\ images of the
50\arcsec$\times$50\arcsec\ fields of all 21 quasars.  The images of
the 12 detected quasars with $>3$ counts were adaptively smoothed
using the algorithm of Ebeling \et (2006). The images of the four
detected sources with $\leq3$ counts, and the images of the fields of
the five undetected quasars, were not smoothed; for these objects the
raw images are shown. We note that the image of SDSS\,J1532$-$0039 was
merged with an archival 5.09\,ks \chandra\ exposure, centered on the
optical position of the source, that did not provide an
\xray\ detection (Vignali \et 2001); the merged image, which
represents a total exposure time of 14.92\,ks, is shown in
Fig.\,\ref{images1}. Inspection of the \chandra\ images of our bright
quasars does not reveal any excess \xray\ emission in their vicinity
that might be interpreted as gravitationally lensed images (e.g.,
Wyithe \& Loeb 2002; Richards \et 2006) or jets (e.g., Schwartz 2002;
Schwartz \& Virani 2004).

With the exception of SDSS\,J0005$-$0006, we did not find any
significant excess of companions with respect to the cumulative number
counts from \xray\ surveys (e.g., Bauer \et 2004) in the
$\approx$300$\times$300\,kpc$^2$ projected regions centered on the
quasar positions; this is in agreement with previous \xray\ searches
around high-$z$ quasars (e.g., Vignali \et 2005 and references
therein). A bright \xray\ source, with $\sim$60 full-band counts, is
apparent in the \chandra\ image of SDSS\,J0005$-$0006 with an angular
distance of $\sim$20\arcsec\ from the target
(Fig.\,\ref{images1}). This source is classified as a galaxy in the
SDSS DR4 catalog (Adelman-McCarthy \et 2006), but due to its faintness
($i=22.1$) an SDSS spectrum is not available. The \xray-to-optical
flux ratio of the source is just $\sim$3 times lower than the
relatively high \xray-to-optical flux ratio of SDSS\,J0005$-$0006 (see
Table\,\ref{properties}, placed at the end of the paper), and based
upon photometry using co-added SDSS images in this area (Jiang \et
2006) its photometric redshift is $z=0.575\pm0.175$ with an integrated
probability of 63\% (Weinstein \et 2004). Given this redshift we
calculate an \xray\ luminosity in the rest-frame \hbox{2--10\,keV}
band of $\approx$10$^{43}$\,erg\,s$^{-1}$ for this AGN candidate (the
luminosity was obtained in the same way as for the quasars in our
sample; see below).

Finally, we searched for rapid variability within the \chandra\
observations of our detected quasars by applying a Kolmogorov-Smirnov
test to the photon arrival times, but no significant flux variations
were detected. This is not surprising given the combination of the
relatively short observations (\ltsim1\,hr in the rest-frame) and the
small number of detected photons from most of the sources.

\setcounter{table}{3}

\begin{figure*}
\plotone{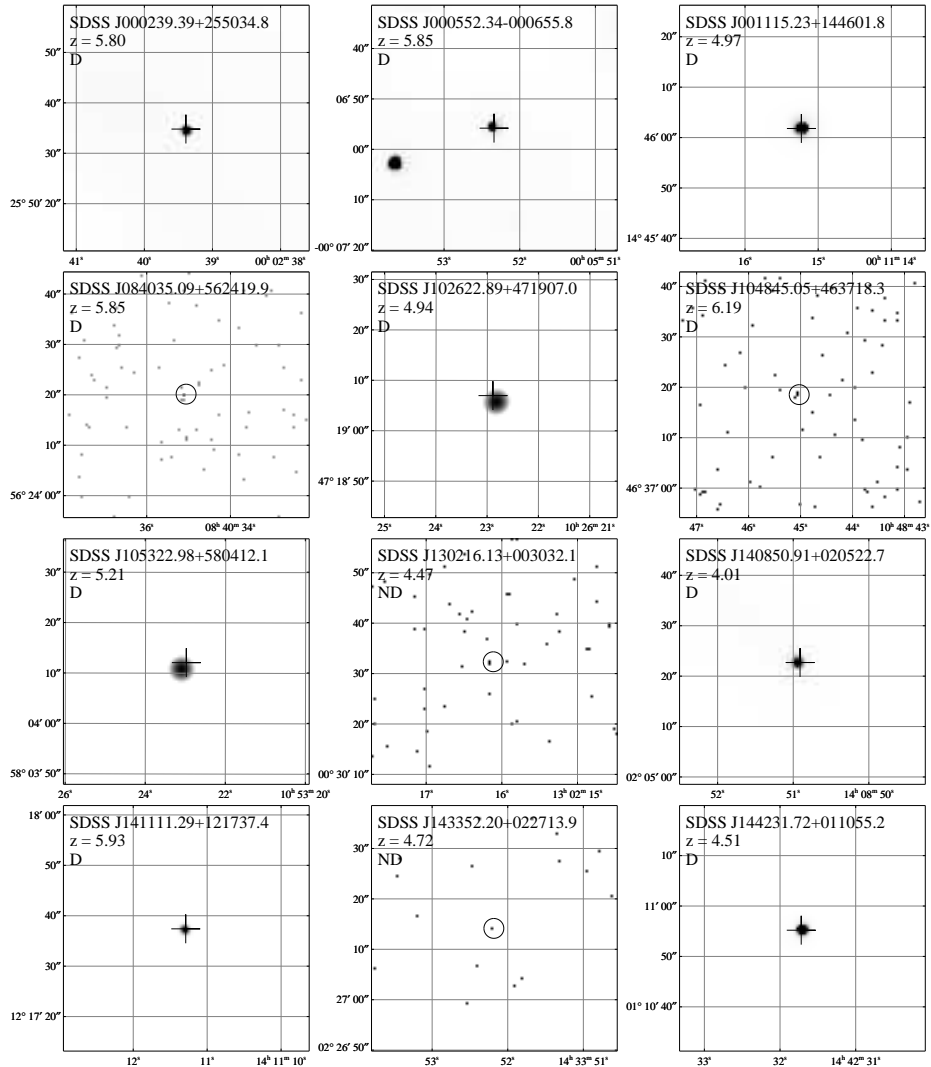}
\vskip -6.0cm
\caption{Full-band (0.5--8\,keV) \chandra\ images of our quasars. Each
  panel spans 50\arcsec$\times$50\arcsec\ on the sky; North is up, and
  East is to the left. The images of the sources with $>3$ counts have
  been adaptively smoothed at the 2$\sigma$ level; the optical
  positions of the quasars are marked by crosses. For sources with
  $\leq3$ counts, the raw (unsmoothed) images and
  \hbox{2\arcsec--radius} circles around the optical positions of the
  quasars are shown. A `D' (`ND') label in each panel indicates an
  \xray\ detection (non-detection) of the source. Note that the image
  of SDSS\,J1532$-$0039 is a combination of our new Cycle~6
  observation and an archival 5.09\,ks \chandra\ exposure; the total
  exposure time for this image is 14.92\,ks. The larger offsets
  between the optical and \xray\ positions of the serendipitous
  sources, SDSS\,J1026$+$4719 and SDSS\,J1053$+$5804, are due to their
  large off-axis angles (see \S\,\ref{observations}).}
\label{images1}
\end{figure*}

\begin{figure*}
\figurenum{\ref{images1}}
\plotone{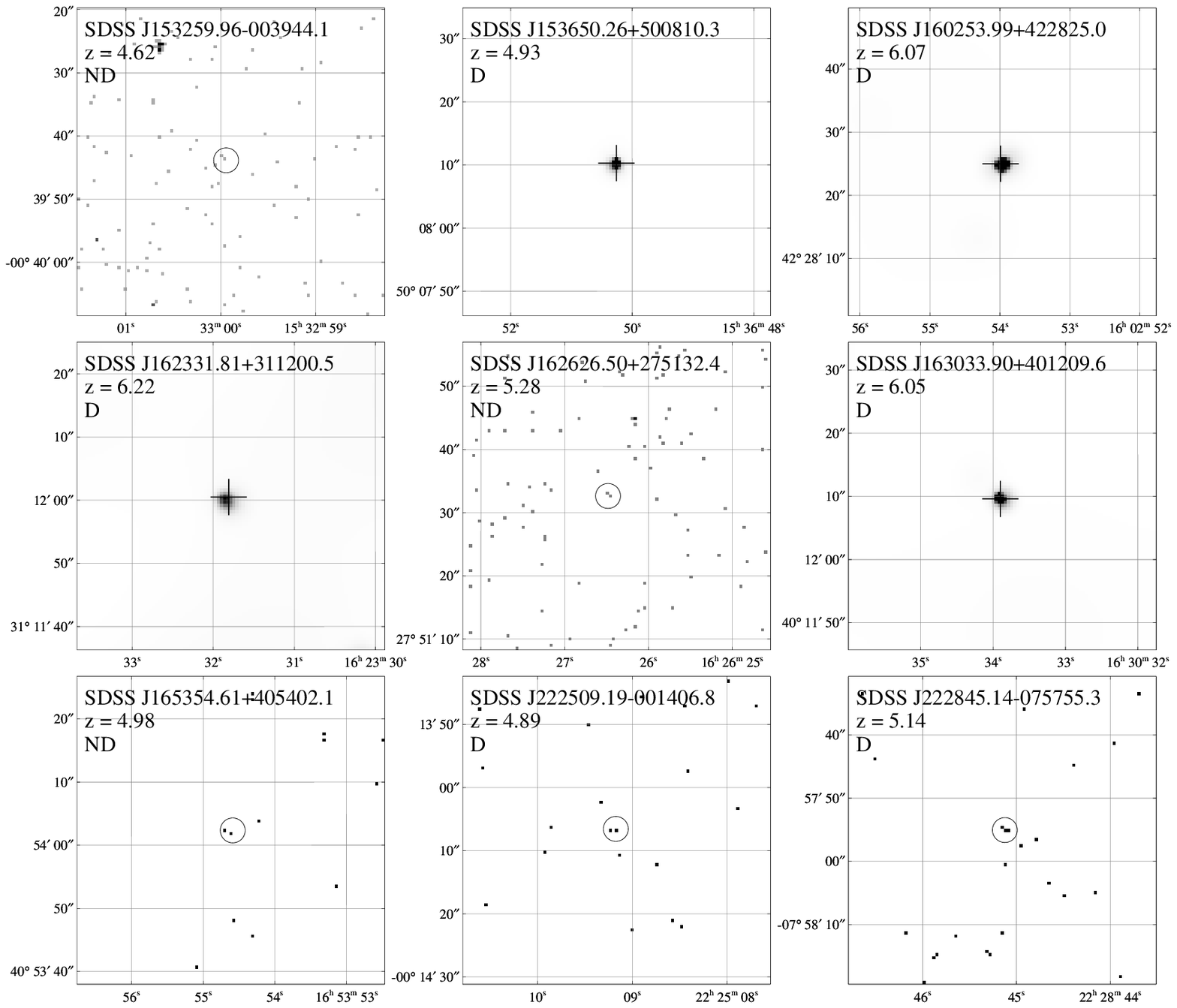}
\vskip -9.5cm
\caption{{\it Continued}}
\end{figure*}

\section{X-ray, Optical, and Radio Properties of the Sample}
\label{results}

The main \xray, optical, and radio properties of our sample are listed
in Table\,\ref{properties}: \\
{\sl Column (1)}. --- The SDSS\,J2000.0 quasar coordinates, accurate to
$\sim0.1$\arcsec. \\
{\sl Column (2)}. --- The Galactic column density in units of
\hbox{10$^{20}$\,cm$^{-2}$}, taken from Dickey \& Lockman (1990) and
obtained with the HEASARC
\nh\
tool.\footnote{http://heasarc.gsfc.nasa.gov/cgi-bin/Tools/w3nh/w3nh.pl}
\\
{\sl Column (3)}. --- The monochromatic $AB$ magnitude at a rest-frame
wavelength of
1450\,\AA\
(\hbox{$AB_{1450}=-2.5\log~f_{1450~\mbox{\scriptsize\AA}}-48.6$}; Oke
\& Gunn 1983). All the magnitudes have been corrected for Galactic
extinction (Schlegel, Finkbeiner, \& Davis 1998). In cases where the
magnitudes were obtained from the SDSS spectra, a fiber light-loss
correction has been applied; this was calculated as the difference
between the synthetic $i$ magnitude (i.e., the integrated flux across
the $i$ bandpass in the SDSS spectrum) and the photometric SDSS $i$
magnitude, assuming no flux variation between the photometric and
spectroscopic epochs. \\
{\sl Column (4)}. --- The absolute $B$-band magnitude, computed
assuming a \hbox{UV--optical} power-law slope of $\alpha=-0.5$
\hbox{($f_{\nu}\propto \nu^{\alpha}$}; Vanden Berk et al. 2001). \\
{\sl Columns (5) and (6)}. --- The flux density and luminosity at a
rest-frame wavelength of 2500\,\AA, computed from the magnitudes in
Column 3, assuming a \hbox{UV--optical} power-law slope of
$\alpha=-0.5$. \\
{\sl Columns (7) and (8)}. --- The count rate in the observed-frame
\hbox{0.5--2\,keV} band and the corresponding flux, corrected for
Galactic absorption and the quantum-efficiency decay of \chandra\ ACIS
at low energy (for the Cycle~4~and~6 sources). The fluxes have been
calculated using {\sc pimms}, assuming a power-law model with
$\Gamma=2.0$, which is a typical photon index for luminous AGN with
$0\ltsim z\ltsim6$ (e.g., Reeves \& Turner 2000; Page \et 2005;
Piconcelli \et 2005; Shemmer \et 2005; Vignali \et 2005; see also
\S\,\ref{xray_properties}). \\
{\sl Columns (9) and (10)}. --- The flux density and luminosity at a
rest-frame energy of 2\,keV, computed assuming $\Gamma=2.0$ and
corrected for the quantum-efficiency decay of \chandra\ ACIS at low
energy (for the Cycle~4~and~6 sources). \\
{\sl Column (11)}. --- The luminosity in the rest-frame
\hbox{2--10\,keV} band. \\
{\sl Column (12)}. --- The optical--X-ray power-law slope, \aox,
defined as:

\begin{equation}
\alpha_{\rm ox}=\frac{\log(f_{\rm
    2\,keV}/f_{2500\mbox{\rm\,\scriptsize\AA}})} {\log(\nu_{\rm
    2\,keV}/\nu_{2500\mbox{\rm\,\scriptsize\AA}})}
\label{eq:aox}
\end{equation}
where $f_{\rm 2\,keV}$ and $f_{2500\mbox{\rm\,\scriptsize\AA}}$ are the
flux densities at rest-frame 2\,keV and 2500\,\AA, respectively. The
errors on \aox\ were computed following the ``numerical method''
described in \S\,1.7.3 of Lyons (1991), taking into account the
uncertainties in the \xray\ count rates and photon indices
($\Gamma=2.0\pm0.1$), and the effects of possible changes in the
\hbox{UV--optical} slope (from $\alpha=-0.5$ to $\alpha=-0.79$). \\
{\sl Column (13)}. --- The difference between the measured \aox\ (from
column\,12) and the predicted \aox, given the UV luminosity from
column\,6, based on the \aox--$L_{\nu}(2500\,\mbox{\AA})$ relation
(given as Eq.\,2 of Steffen \et 2006; see also Strateva \et 2005). The
statistical significance of this difference is also given in units of
$\sigma$, where $\sigma=0.146$ for \hbox{$31<\log
  L_{\nu}(2500\,\mbox{\AA})<32$}, and $\sigma=0.131$ for
\hbox{$32<\log L_{\nu}(2500\,\mbox{\AA})<33$} (see Table\,5 of Steffen
\et 2006). \\
{\sl Column (14)}. --- The radio-loudness parameter (e.g., Kellermann
\et 1989), defined as \hbox{$R=f_{\rm 5\,GHz}/f_{\rm
    4400\,\mbox{\scriptsize\AA}}$} (the ratio of flux densities at
rest-frame frequencies). The flux density at a rest-frame frequency of
5\,GHz was computed from either the FIRST (Becker, White, \& Helfand
1995) or NVSS (Condon et al. 1998) flux densities at an observed-frame
frequency of 1.4\,GHz, assuming a radio power-law slope of
$\alpha=-0.8$. Upper limits on $R$ are at the 3$\sigma$ level, since
the positions of all our sources are known a priori. The flux
densities at rest frame 4400\,\AA\ were computed from the $AB_{1450}$
magnitudes assuming a \hbox{UV--optical} power-law slope of
$\alpha=-0.5$. Typical radio-loudness values are $>100$ for RLQs and
$<10$ for radio-quiet quasars (RQQs). Most of our sources are
RQQs. One source, SDSS\,J0011$+$1446, is a RLQ with $R=136$, and
another source, SDSS\,J1442$+$0110, is radio moderate with
$R=32$. Seven sources have upper limits on $R$ that are not tight
enough to rule out the possibility that they might be radio moderate.

\section{Notes on Individual Objects}
\label{individual}
\begin{figure}
\plotone{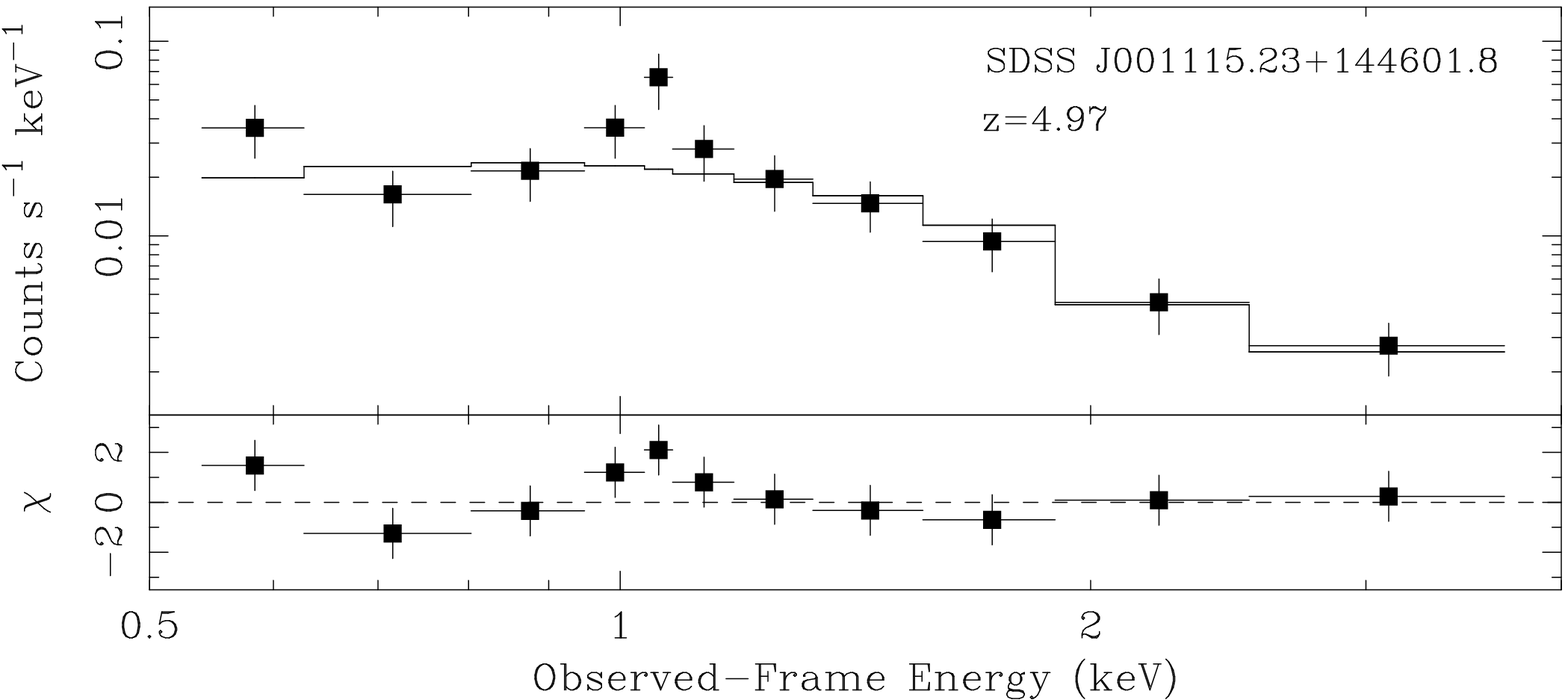}
\caption{\chandra\ spectrum of SDSS\,J0011$+$1446. Filled squares
  represent the binned spectrum, and the solid histogram represents
  the best-fit power-law model. The $\chi$ residuals are in units of
  $\sigma$ with error bars of size 1. Note the 2$\sigma$ residuals at
  $\sim0.6$\,keV and $\sim1.2$\,keV.}
\label{sdssj0011}
\end{figure}

\noindent {\bf SDSS\,J001115.23$+$144601.8} ($z=4.97$): This is the
only RLQ in our sample ($R=136$), and it is one of the most
\xray\ luminous quasars at $z>4$ with $L_{\rm
  2-10\,keV}=3.2\times10^{46}$\,erg\,s$^{-1}$ (cf. Page \et 2005;
Vignali \et 2003b, 2005). The basic \xray, optical, and radio
properties of the source were first presented in Schneider \et
(2005). The \daox$(\sigma)=+0.40\,(3.1)$ of the source indicates that
it significantly deviates from the \aox--$L_{\nu}(2500\,\mbox{\AA})$
relation, being brighter by a factor of $\approx$11 in \hbox{X-rays}
than RQQs of matched UV luminosity. This suggests that the
\xray\ emission from the source is jet dominated. The relatively large
number ($\approx$127) of photons detected from the source also allowed
a basic investigation of its \xray\ spectrum, which was extracted with
the {\sc ciao} task {\sc psextract} using a circular region of
2\arcsec\ radius centered on the \xray\ centroid. The background
region was an annulus with inner and outer radii of 5\arcsec\ and
25\arcsec, respectively. The spectrum was binned in groups of 10
counts per bin. We used {\sc xspec v11.3.2} (Arnaud 1996) to fit the
spectrum with a power-law model and a Galactic absorption component
(Dickey \& Lockman 1990), which was kept fixed during the fit. The
spectrum and its best-fit model and residuals appear in
Fig.\,\ref{sdssj0011}. The best-fit photon index is
$\Gamma=1.75^{+0.37}_{-0.35}$ with $\chi^2=11.02$ for 9 degrees of
freedom (DOF); such a relatively flat power law is typical among RLQs
(e.g., Reeves \& Turner 2000). Adding an intrinsic absorption
component at the redshift of the quasar did not improve the fit, and
in particular did not remove the 2$\sigma$ residuals at $\approx$0.6
and $\approx$1.2\,keV. The residuals at $\approx$1.2\,keV are
consistent in energy with a redshifted \Ka\ line, but we do not expect
such a line in a highly beamed and \xray-luminous source (e.g., Page
\et 2004). The remaining residuals in our spectrum are therefore
difficult to interpret and should be investigated with a higher
signal-to-noise ratio spectrum. \\ \\
{\bf SDSS\,J104845.05$+$463718.3} ($z=6.19$): This is one of two
BALQSOs in our sample. The quasar was discovered by Fan \et (2003),
who detected a hint of a \ion{Si}{4} BAL trough in the optical
spectrum of the source. It was later confirmed as a BALQSO by Maiolino
\et (2004), based on its IR spectrum, and it is the most distant
BALQSO discovered to date. Millimeter observations of this source
detected dust emission and suggested a star-formation rate of
$\sim2000$\,\msun\,yr$^{-1}$ (Bertoldi \et 2003). BALQSOs typically
show soft \xray\ absorption with intrinsic \xray\ column densities in
the range \nh$\approx$10$^{22}$--10$^{24}$\,cm$^{-2}$, rendering them
weak \xray\ sources (e.g., Gallagher \et 1999, 2006). Since we have
accessed highly penetrating \hbox{X-rays} ($\sim$4--70\,keV) from the
source, any \xray\ flux attenuation due to absorption at the quasar
redshift has been mitigated, allowing easier detection. We find that
the source has \aox$=-1.94$, which is consistent with the predicted
value for a non-BAL RQQ with matched UV luminosity (\aox$=-1.72$) at a
level of 1.4$\sigma$ (Table\,\ref{properties}). The \daox$=-0.22$ for
the source is also consistent with the mean difference between the
absorption-corrected \aox\ and the predicted \aox\ values determined
for a sample of 35 BALQSOs (see Figure\,2 of Gallagher \et
2006). \\

\noindent {\bf SDSS\,J130216.13$+$003032.1} ($z=4.47$): This WLQ was
first noted by Anderson \et (2001) to have weak Ly$\alpha$ emission
and an optical spectrum that is similar to the prototype quasar of
this class, SDSS\,J1532$-$0039 (Fan \et 1999). We did not detect this
source in our \chandra\ observation; the upper limit on its \aox,
$-$1.81, is consistent at a level of 1$\sigma$ with the predicted
\aox\ (Table\,\ref{properties}). Thus, the present data cannot prove
that this source is anomalously \xray\ weak. \\ \\
{\bf SDSS\,J140850.91$+$020522.7} ($z=4.01$): This WLQ was first
reported by Schneider \et (2003). Even though it is one of three
quasars in our sample that is radio detected, its radio-to-optical
flux ratio, $R=8.1$, clearly classifies it as radio quiet. It is one
of two WLQs that we detect in the \hbox{X-rays}, and its \aox$=-1.54$
is consistent with the predicted value at a level of 1$\sigma$
(Table\,\ref{properties}). \\ \\
{\bf SDSS\,J144231.72$+$011055.2} ($z=4.51$): This WLQ is radio
moderate with $R=32$. It was first presented by Anderson \et (2001),
who noted its similarity to SDSS\,J1532$-$0039 (Fan \et 1999). The
source is quite \xray\ bright, with $L_{\rm
  2-10\,keV}=2.5\times10^{45}$\,erg\,s$^{-1}$. The relatively flat
\aox$=-1.42$ we find is typical of radio moderate-to-loud quasars, and
it is consistent with the predicted \aox\ for RQQs of matched UV
luminosity at a level of 1.4$\sigma$ (Table\,\ref{properties}); this
indicates a factor of $\sim4$ excess in \xray\ flux compared with the
predicted RQQ value. The weak emission lines may be due to dilution by
a beamed continuum source. The strong \xray\ emission may also suggest
the presence of a moderately beamed \xray\ continuum component, which
is in accordance with the fact that it is only moderately strong in
the radio. \\

\begin{figure}
\plotone{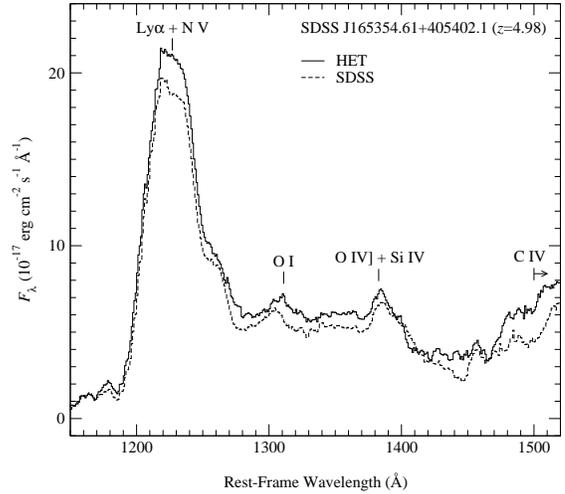}
\caption{HET spectrum of SDSS\,J1653$+$4054 ({\it solid curve})
  superposed on the SDSS spectrum of the source ({\it dashed curve}).
  Both spectra have been smoothed with a 10\,\AA\ boxcar filter. A
  shallow \ion{C}{4} BAL trough is apparent in both spectra at
  $\sim1450$\,\AA. The small differences between the two spectra,
  which are separated by 124\,d in the rest-frame, are probably due to
  systematics of the different instruments and calibration methods
  used.}
\label{HET}
\end{figure}

\noindent {\bf SDSS\,J153259.96$-$003944.1} ($z=4.62$): This is the
prototype high-$z$ WLQ, discovered by Fan \et (1999), who also found a
radio-to-optical spectral index of \aro$>-0.02$\footnote{We adopt the
  definition of \aro\ from Stocke \et (1990); this is equivalent to
  the definition of \aox\ (Eq.\,\ref{eq:aox}), where the 2~keV flux
  density and frequency are replaced by the corresponding values at
  5~GHz.} for the source; this \aro\ lower limit is equivalent to
$R<1.0$ (Table\,\ref{properties}), making this source a RQQ. By
monitoring the source during \hbox{2000--2001}, Stalin \& Srianand
(2005) found that it is optically variable, in the $R$ and $I$ bands,
at a level of $\approx$0.3\,mag; their claim of a $\sim0.9$\,mag
difference between the SDSS photometry obtained in 1998 (Fan \et 1999)
and their 2001 photometry of the source was probably a consequence of
incorrect magnitude conversions between the SDSS and Cousins magnitude
systems.  The source was targeted by \chandra\ in Cycle\,2 but was
undetected in a 5.09\,ks observation (Vignali \et 2001). We targeted
this source with \chandra\ again in Cycle\,6 and observed it for
9.83\,ks. It remains undetected, and the combined exposure time of
14.92\,ks yields \aox$<-1.90$, which is consistent with the predicted
\aox\ at a level of 1.5$\sigma$ (Table\,\ref{properties}). An
additional \chandra\ exposure of $\approx$60\,ks would be required to
determine if this source is notably \xray\ weak (at a 3$\sigma$
level). \\ \\
{\bf SDSS\,J162331.81$+$311200.5} ($z=6.22$): Discovered by Fan \et
(2004), this quasar has the strongest Ly$\alpha$ emission line known
at $z>5$; the total rest-frame equivalent width (REW) of Ly$\alpha$
and \ion{N}{5}\,$\lambda$1240 is \gtsim150\,\AA, compared with a mean
REW(Ly$\alpha$+\ion{N}{5}) of 69.3$\pm$18.0\,\AA\ for $z>4$ quasars
(Fan \et 2001). Otherwise, this source is not particularly optically
luminous; we detected it with \chandra\ and found \aox$=-1.82$, which
is consistent with the predicted value at a 1$\sigma$ level. \\

\begin{deluxetable*}{lcclccl}
\tablecolumns{7} 
\tabletypesize{\small}
\tablewidth{0pc}
\tablecaption{Properties of $z>5$ Radio-Quiet Quasars Included in
  the Joint Fitting}
\tablehead{
\colhead{Object} &
\colhead{} &
\colhead{} &
\colhead{\xray} &
\colhead{Exp. Time\tablenotemark{b}} &
\colhead{Full-Band} &
\colhead{} \\
\colhead{(SDSS\,J)} &
\colhead{$z$} &
\colhead{$N_{\rm H}$\tablenotemark{a}} &
\colhead{Obs. Date} & 
\colhead{(ks)} &
\colhead{Total Counts} &
\colhead{Ref.}
}
\startdata
000239.39$+$255034.8 & 5.80 & 4.09 & 2005 Jan 24 & 5.87 &
{\phn}5.0$^{+3.4}_{-2.2}$ & 1 \\
000552.34$-$000655.8 & 5.85 & 3.14 & 2005 Jul 28 & 16.93 &
{\phn}17.6$^{+5.3}_{-4.1}$ & 1 \\
023137.65$-$072854.5 & 5.41 & 2.90 & 2002 Sep 27 & 4.20 &
{\phn}23.9$^{+6.0}_{-4.9}$ & 2 \\
075618.14$+$410408.6 & 5.11 & 4.76 & 2002 Feb 8 & 7.33 &
{\phn}14.8$^{+4.9}_{-3.8}$ & 2 \\
083643.85$+$005453.3\tablenotemark{c} & 5.77 & 4.40 & 2002 Jan 29 &
5.70 & {\phn}20.7$^{+5.6}_{-4.5}$ & 3 \\ \\
084035.09$+$562419.9 & 5.85 & 4.37 & 2005 Feb 3--4 & 15.84 &
{\phn}3.0$^{+2.9}_{-1.6}$ & 1 \\
103027.10$+$052455.0 & 6.28 & 2.70 & 2002 Jan 29 & 8.10 &
{\phn}5.8$^{+3.6}_{-2.3}$ & 3 \\
105322.98$+$580412.1 & 5.21 & 0.56 & 2001 July 5 & 4.70 &
{\phn}4.3$^{+3.3}_{-2.0}$ & 1 \\
120441.72$-$002149.5 & 5.03 & 2.12 & 2000 Dec 2 & 1.57 &
{\phn}8.0$^{+4.0}_{-2.8}$& 4 \\
130608.26$+$035626.3 & 5.99 & 2.07 & 2002 Jan 29 & 8.20 &
{\phn}16.8$^{+5.2}_{-4.1}$ & 3 \\ \\
141111.29$+$121737.4 & 5.93 & 1.69 & 2005 Mar 17 & 14.27 &
{\phn}12.9$^{+4.7}_{-3.5}$ & 1 \\
160253.98$+$422824.9 & 6.07 & 1.33 & 2005 Oct 29 & 13.20 &
{\phn}26.3$^{+6.2}_{-5.1}$ & 1 \\
162331.81$+$311200.5 & 6.22 & 2.05 & 2004 Dec 29 & 17.21 &
{\phn}5.7$^{+3.5}_{-2.3}$ & 1 \\
163033.90$+$401209.6 & 6.05 & 0.86 & 2005 Nov 4 & 27.39 &
{\phn}17.4$^{+5.3}_{-4.1}$ & 1 \\
222845.14$-$075755.3 & 5.14 & 4.67 & 2003 May 5 & 7.04 &
{\phn}3.0$^{+2.9}_{-1.6}$ & 1
\enddata
\tablenotetext{a}{Neutral Galactic absorption column density in units
  of $10^{20}$\,cm$^{-2}$ taken from Dickey \& Lockman (1990).}
\tablenotetext{b}{The \chandra\ exposure time has been corrected for
  detector dead time.}
\tablenotetext{c}{This source has a radio counterpart in the FIRST
  catalog, with a flux of 1.11$\pm$0.15\,mJy at an observed-frame
  wavelength of 20\,cm; given the optical flux of the source, the
  radio-loudness parameter is $R\ltsim10$, which ranks this quasar, at
  most, as radio moderate (e.g., Brandt \et 2002).}
\tablerefs{ (1) This work; (2) Vignali \et (2003b); (3) Brandt \et
  (2002); (4) Bechtold \et (2003).}
\label{joint_fit_log}
\end{deluxetable*}

\noindent {\bf SDSS\,J165354.61$+$405402.1} ($z=4.98$): Following the
approval of our \chandra\ program, we re-investigated the optical
properties of our approved targets. The SDSS spectrum of this source
(obtained on 2001 June 19) showed a hint of a \ion{C}{4} BAL trough,
which we subsequently confirmed with a 15 minute exposure of the
source, on 2003 July 1, using the Marcarian Low-Resolution
Spectrograph (Hill \et 1998) on the 9\,m Hobby-Eberly Telescope (HET;
Ramsey \et 1998; see Fig.\,\ref{HET}). These spectra suggest that the
quasar is a BALQSO with a \ion{C}{4} absorption trough having
REW$\simeq25$\,\AA\ and an outflow velocity in the range
$1.4\times10^4\ltsim v\ltsim3.1\times10^4$\,\kms. We do not detect
this source in our \chandra\ observation, obtaining \aox$<-1.74$,
which is consistent with the predicted value for a non-BAL RQQ with
matched luminosity at a 0.3$\sigma$ level
(Table\,\ref{properties}). The \daox$<-0.05$ we find
is also consistent with the mean difference between the
absorption-corrected \aox\ and the predicted \aox\ values determined
for a sample of 35 BALQSOs (cf. Figure\,2 of Gallagher \et 2006).

\section{\xray\ Spectral Properties of the Most Distant Quasars}
\label{xray_properties}

Obtaining meaningful \xray\ spectral properties, such as the power-law
photon index and the intrinsic absorption column density, for the
individual sources in this work (with the exception of
SDSS\,J0011$+$1446) is hindered by the small numbers of detected
photons. To date, such measurements have been done for only 10 quasars
at $z>4$ (e.g., Shemmer \et 2005 and references therein). Our
knowledge of the \xray\ spectral properties of quasars at $z>4$
therefore relies mainly on joint spectral fitting of samples of such
sources. For example, Vignali \et (2005) jointly fitted spectra of 48
$z>4$ RQQs at a mean redshift of 4.3; they found a mean photon index
of $\Gamma=1.93^{+0.10}_{-0.09}$ and have constrained the mean
intrinsic neutral absorption column to
\nh\ltsim5$\times$10$^{21}$\,cm$^{-2}$.

We jointly fitted the spectra of all optically selected $z>5$ RQQs,
excluding BALQSOs, which were detected with $>2$ and \ltsim30
full-band counts (the maximum limit prevents one or a few sources from
dominating the average spectrum). This allows investigation of the
mean \xray\ spectral properties of the most distant quasars known,
shedding light on the \xray\ production mechanism and the central
environment of quasars at the end of the reionization epoch. Using
these criteria we found nine suitable sources from this work and six
archival sources, all observed with
\chandra. Table\,\ref{joint_fit_log} lists all 15 sources used in the
joint fitting process along with the total number of counts detected
for each source. The mean (median) redshift of this group of quasars
is 5.72 (5.85), with a total of 185 full-band counts, obtained in
158\,ks.

The \xray\ spectra were extracted with {\sc psextract} using circular
regions of 2\arcsec\ radius centered on the \xray\ centroid of each
source (except for SDSS\,J1053$+$5804, in which a 4\arcsec\ radius was
used due to its off-axis position on the CCD). Although the background
is typically negligible in these \chandra\ observations, we have
extracted background counts using annuli of different sizes (to avoid
contamination from nearby \xray\ sources) centered on each source. We
used {\sc xspec v11.3.2} (Arnaud 1996) to fit the set of 15 unbinned
spectra jointly with the following models: (i) a power-law model and a
Galactic absorption component (Dickey \& Lockman 1990), which was kept
fixed during the fit, and (ii) a model similar to the first with an
added intrinsic (redshifted) neutral-absorption component. All fits
assumed solar abundances (Anders \& Grevesse 1989; Dietrich \et 2003),
used the {\sc phabs} absorption model in {\sc xspec} with the
Balucinska-Church \& McCammon (1992) cross-sections, and used the
$C$-statistic (Cash 1979); this statistical approach allows one to
retain all spectral information and to associate with each quasar its
own Galactic absorption column density and redshift. The errors
associated with the best-fit \xray\ spectral parameters are quoted at
the 90\% confidence level for one parameter of interest ($\Delta
C=2.71$; Avni 1976; Cash 1979).

\begin{deluxetable*}{lcccccc}
\tablecolumns{7} 
\tabletypesize{\scriptsize}
\tablecaption{Best-Fit Parameters from Joint Fitting of $z>5$
  Radio-Quiet Quasars}
\tablehead
{
\colhead{Spectral Model} &
\colhead{Number} &
\colhead{Mean (Median)} &
\colhead{Energy} &
\colhead{\nh} &
\colhead{} &
\colhead{} \\
\colhead{Galactic-Absorbed Power-Law and} &
\colhead{of Sources} &
\colhead{Redshift} &
\colhead{Range\tablenotemark{a}} &
\colhead{(10$^{22}$\,cm$^{-2}$)} &
\colhead{$\Gamma$} &
\colhead{$C$-statistic (bins)}
}
\startdata
\nodata & 15 & 5.72 (5.85) & E & \nodata & $1.95^{+0.27}_{-0.26}$ &
158.6 (175) \\ \\
\nodata & 15 & 5.72 (5.85) & C & \nodata & $1.97^{+0.29}_{-0.28}$ &
154.8 (168) \\ \\
\nodata & 10 & 5.98 (5.96) & E & \nodata & $1.98^{+0.31}_{-0.32}$ &
121.2 (124) \\ \\
Intrinsic Absorption & 15 & 5.72 (5.85) & E & $\le5.98$ &
$1.95^{+0.30}_{-0.26}$ & 158.6 (175) \\ \\
Intrinsic Absorption & 15 & 5.72 (5.85) & C & $\le6.63$ &
$1.97^{+0.30}_{-0.28}$ & 154.8 (168) \\ \\
Intrinsic Absorption & 10 & 5.98 (5.96) & E & $\le8.64$ &
$1.98^{+0.37}_{-0.32}$ & {\phn}121.2 (124)
\enddata
\tablecomments{The joint fitting process was carried out three times
  for each model. In the first run we fitted the entire energy range
  of all $z>5$ quasars, and in the second run we fitted their spectra
  in their common energy range. In the third run we fitted the spectra
  of 10 $z>5.5$ quasars in their entire energy range.}
\tablenotetext{a}{E - entire energy range of all the spectra. C - the
  common rest-frame energy range among all sources
  (\hbox{3.64--48.24\,keV}).}
\label{joint_fit_par}
\end{deluxetable*}

\begin{figure}
\plotone{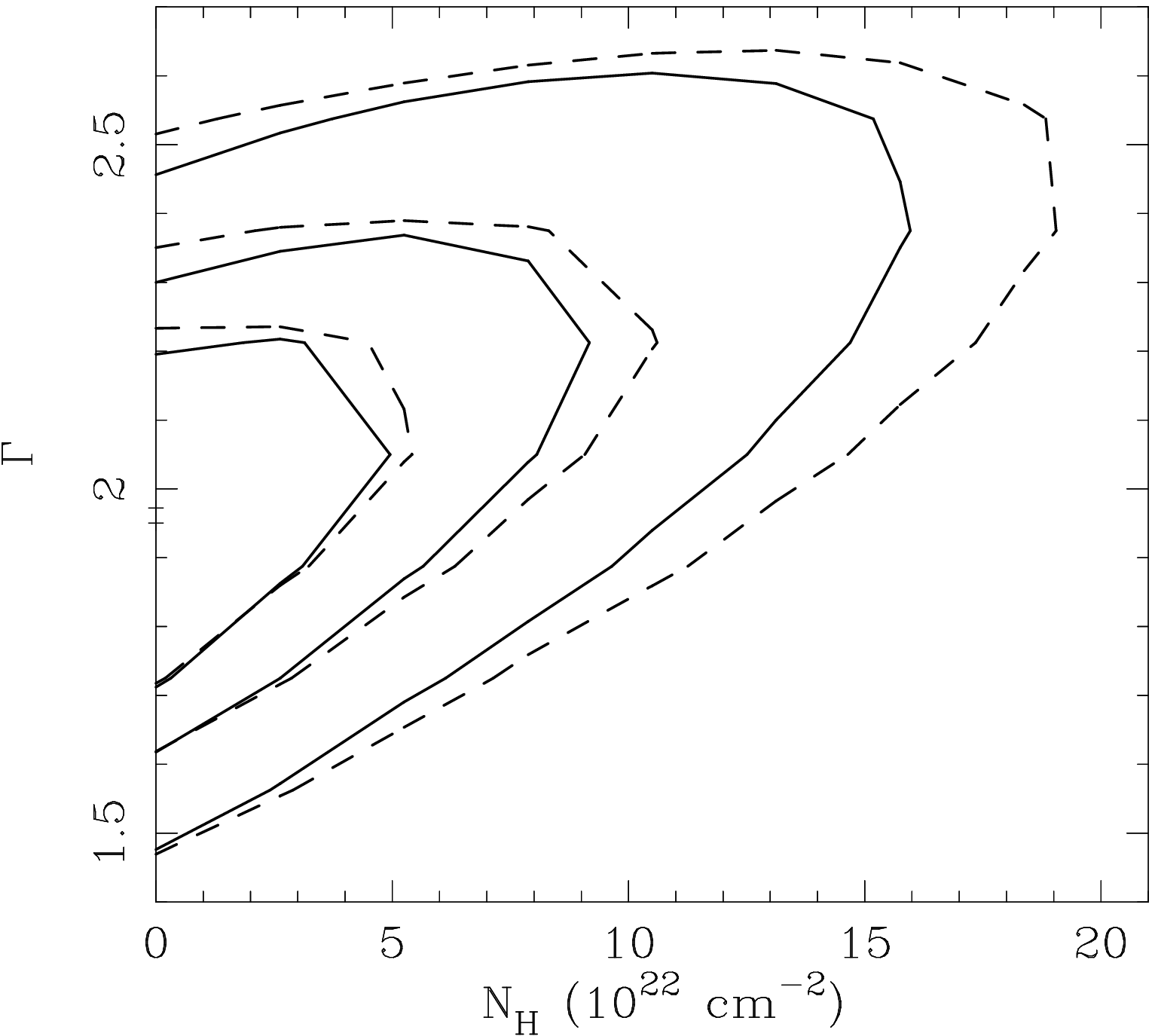}
\caption{68\%, 90\%, and 99\% confidence regions for the photon index
  vs. intrinsic column density derived from joint spectral fitting of
  our sample of 15 RQQs at $z>5$. Solid (dashed) contours refer to the
  entire (common) energy range of the quasars (see
  \S\,\ref{xray_properties} for more details). Note that the markers
  for the best fit parameters in each fit lie on the vertical axis.}
\label{Gamma_NH}
\end{figure}

The joint-fitting process was carried out three times for both models.
In the first run, we included the entire observed-energy range of all
the spectra (\hbox{0.5--8\,keV}). In the second run we included only
the common rest-frame energy range of all 15 sources: 3.64\ltsim
$E_{\rm rest}$\ltsim48.24\,keV (corresponding to \hbox{0.5--8\,keV} in
the observed frame). In the third run we fitted the entire
observed-energy range of ten (out of 15) quasars with $z>5.5$ (the
common rest-frame energy range for these quasars is almost identical
to the entire observed-energy range due to their very narrow redshift
distribution). Table\,\ref{joint_fit_par} lists the best-fit
parameters from the joint-fitting process, and a contour plot of the
$\Gamma$-\nh\ parameter space is shown in Fig.\,\ref{Gamma_NH}. We
find that, in the case of the entire energy range, the mean
\xray\ power-law photon index of $z>5$ RQQs is
$\Gamma=1.95^{+0.30}_{-0.26}$, and their mean neutral intrinsic
absorption column density is \nh\ltsim6$\times$10$^{22}$\,cm$^{-2}$.
These results are consistent with those based on the common energy
range and on the sub-group of $z>5.5$ quasars due to the relatively
narrow redshift range of the sample and the fact that most of the
quasars are at $z>5.5$.

The mean power-law photon index of optically selected $z>5$ RQQs is
consistent with the mean photon index,
\hbox{$\Gamma=1.93^{+0.10}_{-0.09}$}, obtained by jointly fitting
spectra of 48 RQQs at $\left<z\right>=4.3$ (Vignali \et 2005), and
with the mean photon index, $\Gamma=1.97^{+0.06}_{-0.04}$, obtained by
jointly fitting high-quality spectra of eight RQQs at
$\left<z\right>=4.8$ (Shemmer \et 2005). The mean photon index we
obtain for $z>5$ RQQs is also consistent with photon-index
measurements for individual RQQs at lower redshifts (e.g., Reeves \&
Turner 2000; Vignali \et 2001; Page \et 2005; Piconcelli \et
2005). Even though the upper limit we obtained on the mean neutral
intrinsic absorption column in $z>5$ RQQs,
\nh\ltsim6$\times$10$^{22}$\,cm$^{-2}$, is about an order of magnitude
weaker than previous \nh\ upper limits obtained for $z>4$ RQQs with
similar methods (\nh\ltsim5$\times$10$^{21}$\,cm$^{-2}$ at
$\left<z\right>=4.3$; Vignali \et 2005;
\nh\ltsim3$\times10^{21}$\,cm$^{-2}$ at $\left<z\right>=4.8$; Shemmer
\et 2005), it implies that RQQs were not heavily absorbed when the age
of the Universe was \ltsim1\,Gyr.

\begin{figure*}
\plotone{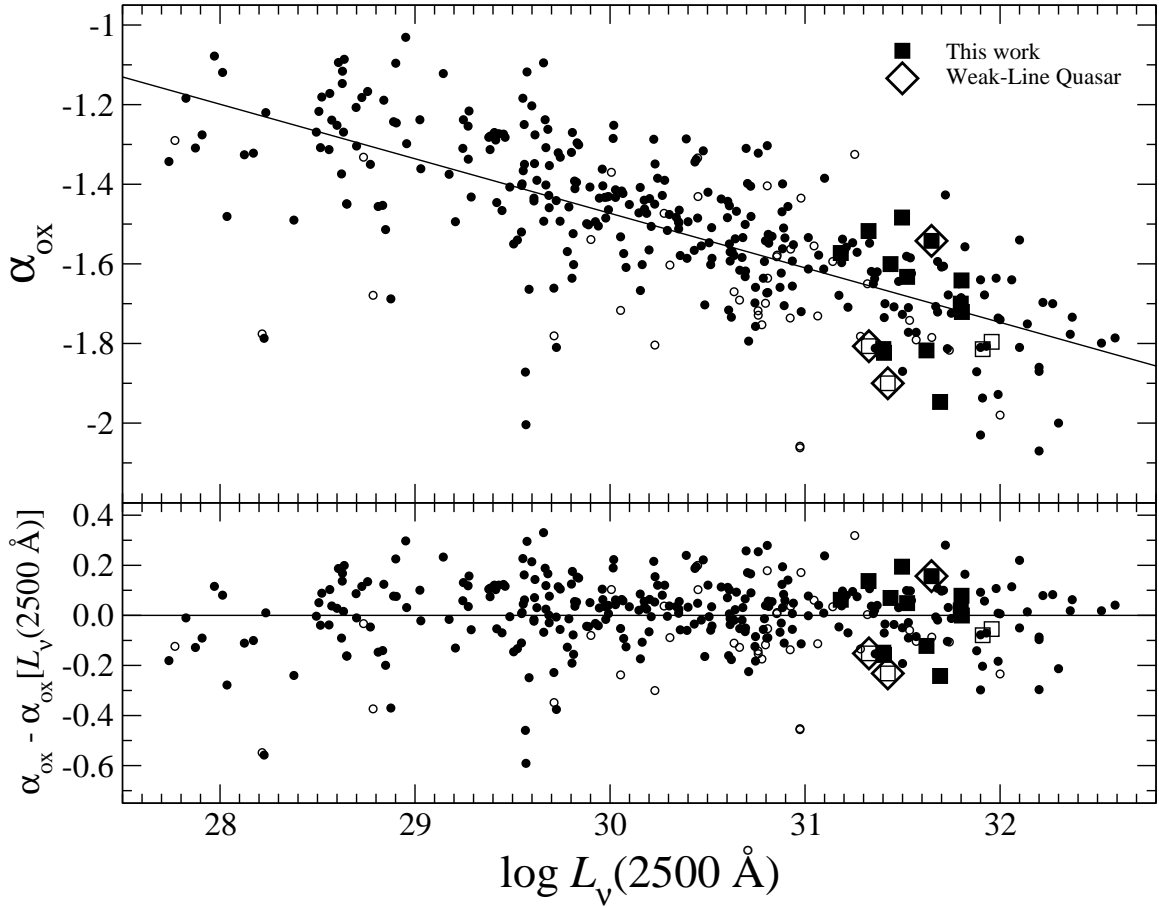}
\caption{The \aox--$L_{\nu}$(2500\,\AA) relationship for RQQs,
  excluding known BALQSOs, adapted from Steffen \et (2006). Open
  symbols indicate \aox\ upper limits. Our sources are marked with
  squares, and enclosing diamonds mark the radio-quiet WLQs in our
  sample (note that the radio-moderate WLQ, SDSS\,J1442$+$0110, is not
  included in the diagram). The Steffen \et (2006) best-fit line is
  shown in the top panel, and the bottom panel shows the
  \aox\ residuals (\daox) from that best-fit line.}
\label{L_2500_a_ox}
\end{figure*}

\begin{figure}
\plotone{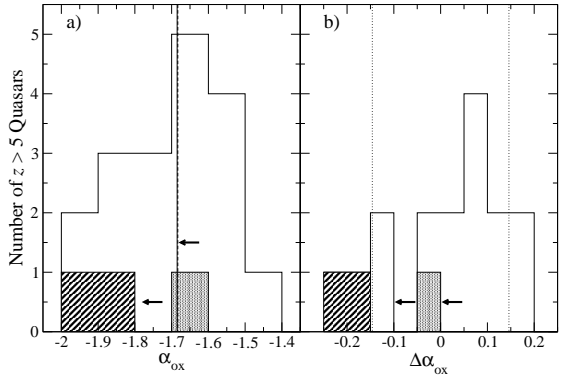}
\caption{Histograms of ({\it a}) \aox\ and ({\it b}) \daox\ for all
  optically selected $z>5$ quasars from this work and the
  literature. Radio-loud quasars are shaded, and BALQSOs are
  hatched. Upper limits are marked by arrows. The solid (dashed) line
  in panel ({\it a}) marks the mean measured (predicted) \aox\ of
  optically selected $z>5$ RQQs, excluding the BALQSOs (note that the
  two lines almost overlap). Dotted lines in panel ({\it b}) mark the
  1$\sigma$ range \hbox{($\sigma=\pm0.146$)} for the distribution of
  \daox\ for sources with \hbox{$31<\log
    L_{\nu}(2500\,\mbox{\AA})<32$} from Steffen \et (2006).}
\label{a_ox_hist}
\end{figure}

We have plotted \aox\ for all the RQQs in our sample against their UV
luminosity in Figure\,\ref{L_2500_a_ox}, and included all 333 RQQs
from the Steffen \et (2006) study in this diagram. Our sources, which
comprise most of the highest-redshift quasars known, follow the
correlation between \aox\ and UV luminosity with no detectable
dependence on redshift. Figure\,\ref{a_ox_hist} shows the
\aox\ distribution of 21 optically selected $z>5$ quasars (11 sources
from this work and 10 from the literature) and the \daox\ distribution
given their UV luminosities. We also calculated the mean \aox\ of the
18 non-BAL RQQs in that sample using the {\sc ASURV} software package
(Lavalley, Isobe, \& Feigelson 1992) to account for two \aox\ upper
limits, and found $\left<\alpha_{\rm ox}\right>=-1.69\pm0.03$ (where
the quoted error is the error on the mean). This value is consistent
with the mean \aox\ predicted from the Steffen \et (2006) relation,
$\left<\alpha_{\rm ox}\right>=-1.68\pm0.15$ (where the quoted error is
the standard deviation), given the mean UV luminosity of the
sources. The \daox\ distribution of $z>5$ quasars indicates that
\aox\ values for most of the sources are consistent within
$\pm$1$\sigma$ with the predicted values, and none of the sources is
significantly \xray\ weak (Fig.\,\ref{a_ox_hist}b). Our results
strongly support the idea that AGN SEDs have not significantly evolved
out to the highest observable redshifts.

\section{X-ray Clues to the Nature of Weak Emission-Line Quasars}
\label{WLQ}

We have observed four members of the rare class of WLQs with
\chandra. Three of these sources are radio quiet, and one,
SDSS\,J1442$+$0110, is radio moderate. SDSS\,J1408$+$0205 and
SDSS\,J1442$+$0110 are the only \xray-detected WLQs in our sample, and
their \aox\ values, $-$1.54 and $-$1.42, respectively, appear
relatively flat although not exceptional
(Table\,\ref{properties}). Both of these sources are also radio
detected, although SDSS\,J1408$+$0205 is, by definition, a RQQ
($R$=8.1). SDSS\,J004054.65$-$091526.8 ($z$=4.98) is another
radio-quiet \xray-detected WLQ; it was serendipitously detected in an
\xmm\ observation of the galaxy cluster Abell 85, and its basic
\xray\ properties are presented in Schneider \et (2003) who found
\aox=$-$1.55. Our data are therefore suggestive of two groups of WLQs:
(i) radio-quiet WLQs that are also \xray\ weak, including
SDSS\,J1302$+$0030 and SDSS\,J1532$-$0039, and (ii) WLQs that are
mildly \xray\ bright, with a relatively flat \aox. The second group
includes the two WLQs with radio detections (although these are not
RLQs), SDSS\,J1408$+$0205 and SDSS\,J1442$+$0110, and the
radio-undetected source SDSS\,J0040$-$0915.

Since the discovery of the prototype WLQ, SDSS\,J1532$-$0039 (Fan \et
1999), about a dozen WLQs have been discovered at $z>2$ with
REW(Ly$\alpha$)$<5$\,\AA\ (Anderson \et 2001; Schneider \et 2003,
2005; Collinge \et 2005; Fan \et 2006). We are also aware of at least
two quasars at lower redshifts that lack high-ionization emission
lines in their spectra, and whose optical luminosities are comparable
to those of our WLQs; these are PG\,1407$+$265 ($z$=0.94; McDowell \et
1995) and PHL\,1811 ($z$=0.19; Leighly \et 2001) that may also be
members of the WLQ class (see also Hawkins 2004). In contrast with
their lack of Ly$\alpha$ and high-ionization lines, PG\,1407$+$265 and
PHL\,1811 exhibit a strong and broad H$\alpha$ and H$\beta$ emission
line, respectively. Another possible member of the WLQ class,
2QZ\,J215454.3$-$305654 at $z$=0.494, shows only an
[\ion{O}{3}]\,$\lambda5007$ emission line, but since UV spectroscopy
of the source was not yet carried out, it is unknown whether it
displays higher ionization lines (Londish \et 2004).

\subsection{Boosted Continuum vs. Depleted High-Ionization Gas}
\label{boosting}

Some of the plausible explanations for the weakness, or absence, of
high-ionization broad emission lines in WLQs include abnormal
photoionizing continuum or broad emission line region (BELR)
properties, and a low BELR covering factor. For example, such quasars
may have a deficit of line-emitting gas in the vicinity of the central
continuum source since an accretion-disk wind, proposed as a
production site for some of the BELR lines (e.g., Murray \& Chiang
1997), may not form, perhaps due to exceptionally high Eddington
ratios (e.g., Leighly, Halpern, \& Jenkins 2004), but see also
Nicastro (2000). As a result of the high accretion rate, the continuum
is expected to be correspondingly softer, thus suppressing
high-ionization emission lines, but having little effect on the
low-ionization species. A softer (\xray) and bluer
(\hbox{UV--optical}) continuum, as well as H$\beta$ and \ion{Fe}{2}
blends are indeed observed in PHL\,1811 (Leighly \et 2004), although
PG\,1407$+$265 displays a `normal' quasar continuum (McDowell \et
1995).

Many featureless-spectrum AGNs are identified as BL Lacertae objects
(hereafter BL Lacs) whose emission lines are diluted by relativistic
beaming of their optical continua (e.g., Urry \& Padovani 1995). Based
on their weak-lined spectra alone, a BL Lac classification is another
plausible explanation for the nature of WLQs. Most BL Lacs are
low-to-moderate luminosity AGNs, discovered at $z$\ltsim1 (e.g.,
Perlman \et 2001 and references therein). By virtue of their weak-line
spectra, the redshifts of BL Lacs are difficult to obtain and can be
highly uncertain. Collinge \et (2005) have found that BL Lac
candidates with no significant proper motions and with either radio or
\xray\ detections tend to lie in specific regions of SDSS optical
color-color diagrams. In particular, their $z<1$ BL Lac candidates
have optical colors that are significantly redder than those of
typical quasars. All of our WLQs have relatively accurate redshift
measurements due to a clear Lyman-break feature, and their optical
spectra resemble those of six of the Collinge \et (2005) BL Lac
candidates at $z>2.15$ (the redshift threshold where Ly$\alpha$ enters
the SDSS spectra).

\begin{figure*}
\plotone{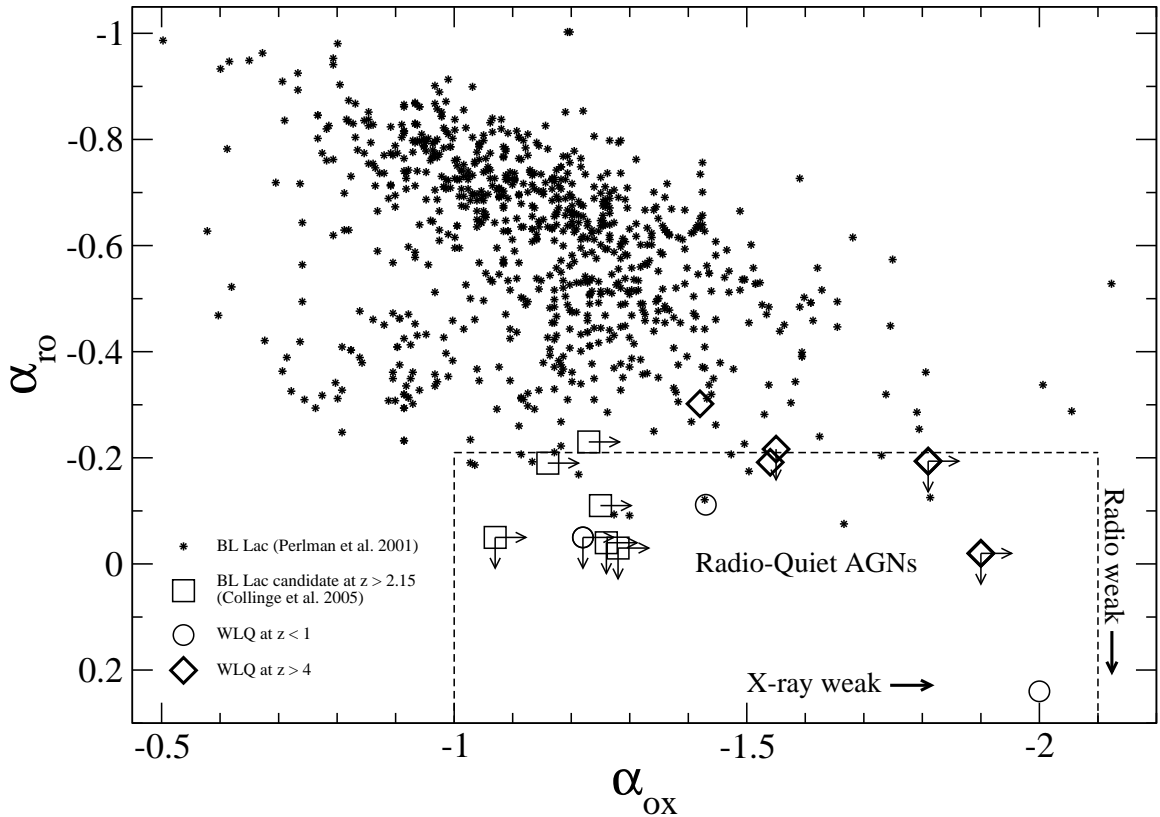}
\caption{The \aro--\aox\ diagram for BL~Lacs, adapted from Perlman \et
  (2001). Stars mark BL Lacs selected from radio and \xray\ surveys
  (Perlman \et 2001, and E. Perlman 2006, private
  communication). Squares mark $z>2.15$ BL Lac candidates from
  Collinge \et (2005), circles mark the $z<1$ WLQs (2QZ\,J2154$-$3056,
  PG\,1407$+$265 and PHL\,1811; the first of these sources is not
  detected in both radio and \hbox{X-rays}, and the latter is the
  radio-weakest source on this diagram), and diamonds mark $z>4$
  WLQs. Right (down) pointing arrows mark \aox\ (\aro) upper
  limits. Dashed lines mark the region containing most radio-quiet
  AGNs. Note that the WLQs and $z>2.15$ BL Lac candidates lie among
  known radio-weak BL Lacs, and that most sources from these three
  groups overlap with radio-quiet AGNs.}
\label{a_ro_a_ox}
\end{figure*}

Most known BL Lacs are strong radio and \xray\ sources (relative to
their optical emission), but the radio/\xray-weak region in the BL Lac
\aro--\aox\ ``color-color'' diagram is being increasingly populated by
deep radio and \xray\ surveys (e.g., Perlman \et 2001 and references
therein). In Figure\,\ref{a_ro_a_ox} we plot the \aro--\aox\ diagram
for all the BL~Lacs from Perlman \et (2001) and E. Perlman (2006),
private communication; we also add to the plot the $z>2.15$ BL~Lac
candidates from Collinge \et (2005), the three $z<1$ WLQs
(2QZ\,J2154$-$3056, PG\,1407$+$265, and PHL\,1811), and the five $z>4$
WLQs discussed in this work. The region occupied by typical
radio-quiet AGNs is also marked in Fig.\,\ref{a_ro_a_ox} (compare with
Fig.\,\ref{L_2500_a_ox}). Apparently, the two radio- and
\xray-detected $z>4$ WLQs (from our second group of WLQs),
SDSS\,J1408$+$0205 and SDSS\,J1442$+$0110, are similar to
(\xray\ selected) radio-weak BL~Lacs. The two radio- and
\xray-undetected $z>4$ WLQs (from our second group of WLQs) and
PHL\,1811 might appear to constitute the tail of the radio/\xray-weak
BL~Lac population, but their radio and \xray\ properties are also
consistent with those of typical RQQs. According to
Fig.\,\ref{a_ro_a_ox}, most of the $z>2.15$ BL~Lac candidates from
Collinge \et (2005) are consistent with the radio and
\xray\ properties of both radio-weak BL~Lacs and typical RQQs,
although the loose constraints on their \aox\ values (based on {\sl
  ROSAT} measurements) prevent a proper distinction between BL~Lacs
and RQQs, as well as a comparison with the WLQs.

Since most BL~Lacs at $z$\ltsim1 are strong radio sources, it is
interesting to note that most WLQs, and the Collinge \et (2005)
$z>2.15$ BL~Lac candidates, are radio weak. The lack of radio-loud
(\aro$<-0.4$, equivalent to $R>100$) WLQs might perhaps be attributed
to the fact that they are all highly optically luminous sources. We
investigated this possibility by examining the luminosities, \aro, and
\aox\ values of samples of lower-redshift ($z$\ltsim1) BL~Lacs. While
BL~Lacs from the EMSS sample (e.g., Rector \et 2000) tend to be more
radio/\xray-weak at high optical luminosities, the BL~Lacs from the
larger RGB sample (Laurent-Muehleisen \et 1999) do not show this
trend, and weak (and strong) radio sources are found at all
luminosities. However, this is not sufficient to rule out a possible
luminosity effect, since even the most luminous BL~Lacs considered are
two orders of magnitude less optically luminous than our WLQs, thus
preventing a straightforward comparison of the radio properties
between the two classes of sources. The lack of highly optically
luminous BL~Lacs may be a consequence of their selection, which is
performed mainly with radio and \xray\ surveys. Since it is possible
that these selection methods have allowed highly optically luminous
BL~Lacs to escape detection, we checked for a
radio-loudness--luminosity relation among the Collinge \et (2005)
optically-selected sample of low-redshift BL~Lac candidates, but no
clear trend has been found.

Two other well-known characteristics of BL~Lacs are rapid and
large-amplitude optical flux variations, and high levels (up to 40\%)
of optical polarization (e.g., Jannuzi, Smith, \& Elston 1994).
Stalin \& Srianand (2005) have detected optical flux variations at a
level of $\approx$0.3\,mag in the prototype WLQ, SDSS\,J1532$-$0039,
on an observed-frame time baseline of 450\,d (80\,d in the
rest-frame), although this is somewhat lower than, and not as fast as,
the level displayed by typical BL~Lacs (e.g., Ghosh \et 2000). Fan \et
(1999) have constrained the optical polarization level of
SDSS\,J1532$-$0039 to be $\leq4$\%. This is consistent with
single-epoch measurements of polarization levels in \xray-selected
BL~Lacs, which are typically lower than the polarization levels
observed in radio-selected BL~Lacs (e.g., Jannuzi \et 1994;
Visvanathan \& Wills 1998). Therefore, low optical polarization levels
in WLQs might be expected due to their radio weakness, regardless of
whether they are BL~Lacs or not.

While the nature of WLQs remains mysterious and apparently there is
significant variety among this class of sources, a `boosted continuum'
or a `suppression of high-ionization BELR lines' remain the most
likely scenarios. The significant overlap in the radio and
\xray\ properties between different AGN classes (i.e., BL~Lacs and
radio-quiet AGNs) as well as the multiple selection biases, as
depicted in the \aro--\aox\ diagram (Fig.\,\ref{a_ro_a_ox}), might
cause physically distinct AGN populations to blend together, and thus
prevent a clear conclusion for the nature of WLQs. This calls for a
more thorough and systematic investigation of the `continuum boosting'
versus `high-ionization emission-line suppression' scenarios via more
observations such as those described below, and detailed
photoionization modeling. While relativistic boosting of the continuum
may be consistent with the optical spectra, radio fluxes, and
\xray\ fluxes of the sources from our second group of WLQs as well as
with some of the $z>2.15$ BL~Lac candidates of Collinge \et (2005), it
cannot hold for the two $z<1$ WLQs that display Balmer emission
lines. It is also hard to accept a boosting scenario for our first
group of WLQs and most of the $z>2.15$ BL~Lac candidates of Collinge
\et (2005), due to the weakness of their radio and \xray\ emission. In
addition, the fact that radio-loud WLQs (with $R>100$) are missing is
naturally explained by the `line suppression' scenario, which does not
require that the quasars be radio-loud. The following observations for
all $z>4$ WLQs and the $z>2.15$ BL~Lac candidates of Collinge \et
(2005) are required to enable a clearer distinction between the two
proposed scenarios:

\begin{enumerate}
\item{Sensitive near-IR spectroscopy to reveal any low-ionization
  emission lines (e.g., Balmer or \ion{Fe}{2} lines) and determine
  whether the \hbox{UV--optical} continua are typical of RQQs. Clear
  detection of low-ionization lines, and/or a typical (or bluer)
  continuum will support the `high-ionization line suppression'
  scenario, while the non-detection of such lines and redder continua
  will support the BL~Lac interpretation.}
\item{{\sl Spitzer} near--mid-IR observations (using {\sl IRAC} and
  {\sl MIPS}) may determine whether the SED is consistent with a
  typical AGN IR `bump' or with a beamed, `bump-less', continuum.}
\item{Optical--near-IR photometric monitoring to detect rapid and
  large-amplitude flux variations can check for a boosted continuum.
  \\ \\ }
\end{enumerate}

\subsection{A Macro- or Micro-Lensing Interpretation?}
\label{lensing}

The absence of prominent optical emission lines in luminous,
high-redshift quasars has raised several other possible explanations
for their peculiar spectra. One of these is that high-redshift WLQs
are not quasars at all, but rather normal galaxies (e.g., Ivison \et
2005). Their measured luminosities are much larger than even the most
luminous galaxies, but could perhaps be explained by magnification due
to gravitational lensing. Our \xray\ observations rule out that
possibility, at least for the cases of the three \xray-detected
high-redshift WLQs that show AGN-like \xray\ properties, i.e., typical
AGN optical-to-X-ray flux ratios (see the \aox\ values in
Table\,\ref{properties} and in Schneider \et 2003; normal galaxies
typically have \aox$\ll-2$).

Another possibility is that the continuum source in high-redshift WLQs
is amplified relative to the emission from the, much larger, broad
emission-line region by gravitational microlensing, since the degree
of amplification is inversely proportional to the size of the emitting
region (e.g., Paczynski 1986; Schneider \& Weiss 1987; Ostriker \&
Vietri 1990). Such microlensing events are expected to be both
relatively short (months--years) and rare, and they might perhaps
qualitatively explain the rarity of high-redshift WLQs among the
quasar population (e.g., Wambsganss, Schneider, \& Paczynski
1990). Testing this scenario would require spectroscopic monitoring of
our high-redshift WLQs to search for the reappearance of the emission
lines. However, the SDSS images of our WLQs, and the \xray\ images of
our \xray-detected WLQs, do not show evidence for multiple images or
lensing galaxies in their vicinity, and therefore a microlensing
interpretation is unlikely (we also note that our WLQs were not
targeted with the {\sl Hubble Space Telescope} by Richards \et 2006).

\section{Summary}
\label{summary}

We present new \chandra\ observations of 21 $z>4$ quasars, including
most of the highest-redshift quasars known as well as four members of
the WLQ class. We have doubled the number of \xray-detected quasars at
$z>5$, allowing investigation of the \xray\ spectral properties of
AGNs at the end of the reionization epoch. Our main results can be
summarized as follows:
\begin{enumerate}
\item{The mean power-law \xray\ photon index of $z>5$ RQQs is
  $\Gamma$=1.95$^{+0.30}_{-0.26}$, and it is consistent with photon
  indices observed for RQQ samples at lower redshifts.}
\item{The upper limit on the mean intrinsic-absorption column density
  of $z>5$ RQQs is \nh\ltsim6$\times$10$^{22}$\,cm$^{-2}$, implying
  that RQQs at the dawn of the modern Universe were not heavily
  absorbed.}
\item{The \aox\ values for individual RQQs in our sample, as well as
  the mean \aox\ of $z>5$ RQQs, excluding BALQSOs, are consistent with
  the values predicted from their UV luminosities. This strengthens
  previous findings that AGN SEDs have not significantly evolved over
  cosmic time.}
\item{We detected two of our $z>4$ WLQs in \hbox{X-rays}, and
  distinguished two types of WLQs: radio-quiet WLQs which are also
  \xray\ weak, and radio-quiet--moderate WLQs (i.e., with $R<100$)
  that are mildly \xray\ bright. Based on single-epoch optical spectra
  as well as \xray\ and radio fluxes of these sources, we discuss
  several possible interpretations for the weakness of the
  broad-emission lines, and in particular, continuum boosting versus
  high-ionization emission-line suppression, which are the more likely
  scenarios. We propose sensitive near-IR spectroscopy, near--mid-IR
  photometry, and optical--near-IR monitoring to disentangle these two
  scenarios.}
\end{enumerate}

\acknowledgments

We gratefully acknowledge the financial support of
\chandra\ \xray\ Center grants GO3-4117X and GO5-6094X (O.\,S.,
W.\,N.\,B., D.\,P.\,S), NASA LTSA grant NAG5-13035 (O.\,S.,
W.\,N.\,B., D.\,P.\,S.), and NSF grants AST-0307582 (D.\,P.\,S.),
AST-0307384 (X.\,F.), and AST-0307409 (M.\,A.\,S.). We thank Pat Hall,
Aaron Steffen, and Iskra Strateva for helpful comments and
discussions, and Linhua Jiang for obtaining the photometric data for
SDSS\,J0005$-$0006. We also thank Eric Perlman for kindly providing
the \aro\ and \aox\ data for Fig.\,\ref{a_ro_a_ox}. An anonymous
referee is gratefully acknowledged for a prompt and constructive
report that helped to improve the manuscript. Funding for the creation
and distribution of the SDSS Archive has been provided by the Alfred
P. Sloan Foundation, the Participating Institutions, the National
Aeronautics and Space Administration, the National Science Foundation,
the U.S. Department of Energy, the Japanese Monbukagakusho, and the
Max Planck Society. The SDSS Web site is http://www.sdss.org/. The
SDSS is managed by the Astrophysical Research Consortium (ARC) for the
Participating Institutions. The Participating Institutions are The
University of Chicago, Fermilab, the Institute for Advanced Study, the
Japan Participation Group, The Johns Hopkins University, the Korean
Scientist Group, Los Alamos National Laboratory, the
Max-Planck-Institute for Astronomy (MPIA), the Max-Planck-Institute
for Astrophysics (MPA), New Mexico State University, University of
Pittsburgh, University of Portsmouth, Princeton University, the United
States Naval Observatory, and the University of Washington. The HET is
a joint project of the University of Texas at Austin, the Pennsylvania
State University, Stanford University,
Ludwig-Maximillians-Universit\"at M\"unchen, and
Georg-August-Universit\"at G\"ottingen. The HET is named in honor of
its principal benefactors, William~P.~Hobby and Robert~E.~Eberly.
This research has made use of the NASA/IPAC Extragalactic Database
(NED) which is operated by the Jet Propulsion Laboratory, California
Institute of Technology, under contract with the National Aeronautics
and Space Administration.

\setcounter{table}{2}

\clearpage

\begin{landscape}
\begin{deluxetable}{cccccccccccccc}
\tablecolumns{14}
\tablewidth{0pt}
\tablecaption{Optical, X-ray, and Radio Properties}
\tablehead{
\colhead{} &
\colhead{} &
\colhead{} &
\colhead{} &
\colhead{} &
\colhead{$\log(\nu L_\nu)$} &
\colhead{Count} & 
\colhead{} &
\colhead{} &
\colhead{$\log (\nu{L_\nu})$} &
\colhead{$\log L$} &
\colhead{} &
\colhead{} &
\colhead{} \\
\colhead{Object (SDSS\,J)} &
\colhead{$N_{\rm H}$\tablenotemark{a}} &
\colhead{$AB_{1450}$} &
\colhead{$M_B$} &
\colhead{$f_{2500\,{\rm \AA}}$\tablenotemark{b}} &
\colhead{2500~\AA} &
\colhead{Rate\tablenotemark{c}} &
\colhead{$f_{\rm x}$\tablenotemark{d}} &
\colhead{$f_{2\,\rm keV}$\tablenotemark{e}} &
\colhead{${2\,\rm keV}$} &
\colhead{2--10~keV} &
\colhead{$\alpha_{\rm ox}$} &
\colhead{\daox~$(\sigma)$\tablenotemark{f}} &
\colhead{$R$\tablenotemark{g}} \\
\colhead{(1)} &
\colhead{(2)} &
\colhead{(3)} &
\colhead{(4)} &
\colhead{(5)} &
\colhead{(6)} &
\colhead{(7)} &
\colhead{(8)} &
\colhead{(9)} &
\colhead{(10)} &
\colhead{(11)} &
\colhead{(12)} &
\colhead{(13)} &
\colhead{(14)}
}
\startdata
000239.39$+$255034.8 & 4.09 & 19.0\tablenotemark{h} & $-$28.0 & 11.8 &
46.9 & {\phn}0.85$^{+0.58}_{-0.37}$ & {\phn}3.8$^{+2.5}_{-1.7}$ & 3.86
& {\phn}45.0 & {\phn}45.2 & {\phn}$-$1.72$^{+0.09}_{-0.10}$ &
$-0.00~(0.0)$ & $<26.8$\tablenotemark{o} \\
000552.34$-$000655.8 & 3.14 & 20.2\tablenotemark{h} & $-$26.8 & 3.9 &
46.4 & {\phn}0.97$^{+0.30}_{-0.24}$ & {\phn}4.2$^{+1.3}_{-1.1}$ & 4.29
& {\phn}45.1 & {\phn}45.3 & {\phn}$-$1.52$^{+0.06}_{-0.05}$ &
$+0.14~(0.9)$ & $<51.4$ \\
001115.23$+$144601.8 & 4.28 & 18.0\tablenotemark{i} & $-$28.7 & 29.9 &
47.2 & {\phn}28.12$^{+3.14}_{-2.83}$ & {\phn}97.1$^{+10.8}_{-9.8}$ &
86.49 & {\phn}46.3 & {\phn}46.5 & {\phn}$-$1.36$^{+0.04}_{-0.03}$ &
$+0.40~(3.1)$ & 135.9\tablenotemark{o} \\
084035.09$+$562419.9 & 4.37 & 20.0\tablenotemark{j} & $-$27.0 & 4.6 &
46.5 & {\phn}$0.19^{+0.19}_{-0.10}$ & $0.8^{+0.8}_{-0.4}$ & 0.86 &
{\phn}44.4 & {\phn}44.6 & {\phn}$-1.81^{+0.12}_{-0.12}$ &
$-0.15~(1.0)$ & $<11.8$ \\
102622.89$+$471907.0 & 1.25 & 18.8\tablenotemark{i} & $-$28.0 & 14.9 &
46.9 & {\phn}1.85$^{+1.55}_{-0.91}$\tablenotemark{m} &
{\phn}8.9$^{+7.5}_{-4.4}$ & 7.86 & {\phn}45.2 & {\phn}45.4 &
{\phn}$-1.64$$^{+0.11}_{-0.12}$ & $+0.08~(0.5)$ & $<3.0$ \\
104845.05$+$463718.3 & 1.26 & 19.3\tablenotemark{k} & $-$27.9 & 9.5 &
46.8 & {\phn}0.20$^{+0.20}_{-0.11}$ & {\phn}0.8$^{+0.8}_{-0.4}$ & 0.86
& {\phn}44.4 & {\phn}44.6 & {\phn}$-$1.94$^{+0.12}_{-0.12}$ &
$-0.22~(1.4)$ & $<6.0$ \\
105322.98$+$580412.1 & 0.56 & 19.6\tablenotemark{i} & $-$27.3 & 7.2 &
46.6 & {\phn}0.92$^{+0.78}_{-0.46}$\tablenotemark{m} &
{\phn}4.3$^{+3.7}_{-2.2}$ & 4.02 & {\phn}45.0 & {\phn}45.2 &
{\phn}$-1.63$$^{+0.11}_{-0.12}$ & $+0.05~(0.3)$ & $<7.3$ \\ \\
130216.13$+$003032.1 & 1.57 & 19.8\tablenotemark{i} & $-$26.8 & 5.8 &
46.4 & {\phn}$<0.44$ & $<1.4$ & $<1.14$ & {\phn}$<44.3$ &
{\phn}$<44.5$ & {\phn}$<-1.81$ & $<-0.16~(>1.0)$ & $<8.3$ \\
140850.91$+$020522.7 & 2.57 & 18.8\tablenotemark{i} & $-$27.6 & 14.4 &
46.7 & {\phn}4.36$^{+1.03}_{-0.85}$ & {\phn}18.5$^{+4.4}_{-3.6}$ &
13.84 & {\phn}45.3 & {\phn}45.5 & {\phn}$-$1.54$^{+0.05}_{-0.04}$ &
$+0.16~(1.0)$ & 8.1 \\
141111.29$+$121737.4 & 1.69 & 20.0\tablenotemark{h} & $-$27.1 & 4.9 &
46.5 & {\phn}0.77$^{+0.31}_{-0.23}$ & {\phn}3.2$^{+1.3}_{-1.0}$ & 3.31
& {\phn}45.0 & {\phn}45.2 & {\phn}$-$1.60$^{+0.07}_{-0.07}$ &
$+0.07~(0.5)$ & $<13.5$ \\
143352.20$+$022713.9 & 2.74 & 18.4\tablenotemark{i} & $-$28.3 & 20.5 &
47.0 & {\phn}$<1.37$ & $<4.5$ & $<3.84$ & {\phn}$<44.9$ &
{\phn}$<45.1$ & {\phn}$<-1.81$ & $<-0.08~(>0.5)$ & $<2.4$ \\
144231.72$+$011055.2 & 3.33 & 20.0\tablenotemark{i} & $-$26.6 & 4.7 &
46.3 & {\phn}3.33$^{+0.65}_{-0.55}$ & {\phn}11.2$^{+2.1}_{-1.9}$ &
9.21 & {\phn}45.2 & {\phn}45.4 & {\phn}$-$1.42$^{+0.04}_{-0.04}$ &
$+0.22~(1.4)$ & 31.7 \\
153259.96$-$003944.1 & 6.60 & 19.6\tablenotemark{l} & $-$27.0 & 6.9 &
46.5 & {\phn}$<0.20$\tablenotemark{n} & $<0.9$ & $<0.79$ &
{\phn}$<44.2$ & {\phn}$<44.4$ & {\phn}$<-1.90$ & $<-0.23~(>1.5)$ &
$<1.0$\tablenotemark{p} \\
153650.26$+$500810.3 & 1.57 & 18.8\tablenotemark{i} & $-$28.0 & 14.8 &
46.9 & {\phn}1.51$^{+0.82}_{-0.56}$ & {\phn}6.2$^{+3.4}_{-2.3}$ & 5.51
& {\phn}45.1 & {\phn}45.3 & {\phn}$-$1.70$^{+0.08}_{-0.08}$ &
$+0.02~(0.1)$ & $<3.2$ \\ \\
160253.98$+$422824.9 & 1.33 & 19.9\tablenotemark{h} & $-$27.2 & 5.4 &
46.6 & {\phn}1.71$^{+0.44}_{-0.36}$ & {\phn}7.0$^{+1.8}_{-1.5}$ & 7.39
& {\phn}45.3 & {\phn}45.5 & {\phn}$-$1.48$^{+0.05}_{-0.05}$ &
$+0.20~(1.3)$ & $<10.3$ \\
162331.81$+$311200.5 & 2.05 & 20.1\tablenotemark{h} & $-$27.0 & 4.2 &
46.5 & {\phn}0.17$^{+0.17}_{-0.09}$ & {\phn}0.7$^{+0.7}_{-0.4}$ & 0.75
& {\phn}44.3 & {\phn}44.5 & {\phn}$-$1.82$^{+0.12}_{-0.14}$ &
$-0.16~(1.0)$ & $<12.3$ \\
162626.50$+$275132.4 & 3.27 & 18.5\tablenotemark{i} & $-$28.4 & 19.3 &
47.0 & {\phn}$<1.00$ & $<4.3$ & $<4.03$ & {\phn}$<45.0$ &
{\phn}$<45.2$ & {\phn}$<-1.80$ & $<-0.05~(>0.4)$ & $<2.6$ \\
163033.90$+$401209.6 & 0.86 & 20.6\tablenotemark{k} & $-$26.4 & 2.6 &
46.3 & {\phn}0.50$^{+0.17}_{-0.13}$ & {\phn}2.0$^{+0.7}_{-0.5}$ & 2.11
& {\phn}44.8 & {\phn}45.0 & {\phn}$-$1.57$^{+0.06}_{-0.05}$ &
$+0.06~(0.4)$ & $<19.6$ \\
165354.61$+$405402.1 & 1.94 & 19.3\tablenotemark{i} & $-$27.5 & 9.4 &
46.7 & {\phn}$<0.96$ & $<3.1$ & $<2.77$ & {\phn}$<44.8$ &
{\phn}$<45.0$ & {\phn}$<-1.74$ & $<-0.05~(>0.3)$ & $<5.3$ \\
222509.19$-$001406.8 & 5.03 & 19.2\tablenotemark{i} & $-$27.5 & 10.0 &
46.7 & {\phn}0.58$^{+0.77}_{-0.37}$ & {\phn}2.1$^{+2.7}_{-1.4}$ & 1.84
& {\phn}44.6 & {\phn}44.8 & {\phn}$-$1.82$^{+0.14}_{-0.18}$ &
$-0.12~(0.8)$ & $<4.7$ \\
222845.14$-$075755.3 & 4.67 & 19.1\tablenotemark{i} & $-$27.7 & 10.9 &
46.8 & {\phn}0.28$^{+0.38}_{-0.18}$ & {\phn}1.0$^{+1.3}_{-0.7}$ & 0.92
& {\phn}44.3 & {\phn}44.5 & {\phn}$-$1.95$^{+0.14}_{-0.20}$ &
$-0.24~(1.6)$ & $<4.4$ \\
\enddata
\tablenotetext{a}{Neutral Galactic absorption column density in units
  of $10^{20}$\,cm$^{-2}$ taken from Dickey \& Lockman (1990).}
\tablenotetext{b}{Flux density at rest-frame 2500\,\AA\ in units of
  10$^{-28}$\,erg\,cm$^{-2}$\,s$^{-1}$\,Hz$^{-1}$.}
\tablenotetext{c}{Observed count rate computed in the
  \hbox{0.5--2\,keV} band in units of 10$^{-3}$\,counts\,s$^{-1}$.}
\tablenotetext{d}{Galactic absorption-corrected flux in the observed
  \hbox{0.5--2\,keV} band in units of
  10$^{-15}$\,erg\,cm$^{-2}$\,s$^{-1}$.}
\tablenotetext{e}{Flux density at rest-frame 2\,keV in units of
  10$^{-32}$\,erg\,cm$^{-2}$\,s$^{-1}$\,Hz$^{-1}$.}
\tablenotetext{f}{The difference between measured and predicted
  \aox\ ($\Delta$\aox), and the significance of that difference
  ($\sigma$), based on the Steffen \et (2006)
  \aox--$L_{\nu}(2500\,\mbox{\AA})$ relation.}
\tablenotetext{g}{Unless otherwise noted, radio-to-optical flux ratios
  ($R$) involve radio flux densities at an observed-frame frequency of
  1.4\,GHz taken from the FIRST survey (Becker \et 1995); upper limits
  on $R$ are calculated from the 3$\sigma$ FIRST detection threshold
  at source position.}
\tablenotetext{h}{Taken from Fan \et (2004).}
\tablenotetext{i}{Obtained from the SDSS spectrum, corrected for
  Galactic extinction (Schlegel \et 1998) and fiber light loss (see
  \S\,\ref{results}).}
\tablenotetext{j}{Taken from Fan \et (2006).}
\tablenotetext{k}{Taken from Fan \et (2003).}
\tablenotetext{l}{Estimated as [($z+$0.1)$+$($i-$0.2)]$/$2 using the
  $i$ and $z$ magnitudes in Fan \et (1999).}
\tablenotetext{m}{Count rate corrected for the exposure map due to a
  large offset from the aimpoint (see \S\,\ref{observations}).}
\tablenotetext{n}{Constraints on the count rate and subsequent \xray\
  parameters were obtained by including an additional \chandra\
  exposure (see \S\,\ref{observations}).}
\tablenotetext{o}{Flux density at an observed-frame frequency of
  1.4\,GHz taken from the NVSS survey (Condon \et 1998); upper limits
  are calculated from the NVSS detection threshold of 2.5\,mJy.}
\tablenotetext{p}{Taken from Fan \et (1999).}
\label{properties}
\end{deluxetable}
\clearpage

\end{landscape}

\end{document}